\documentclass[a4paper,11pt]{article}
\pdfoutput=1 

\usepackage{jheppub} 

\reversemarginpar

\usepackage{graphicx}
\usepackage{amsmath}
\usepackage{hyperref}
\usepackage{pdfpages}
\usepackage[section]{placeins}
\usepackage{epstopdf}
\usepackage[table,dvipsnames]{xcolor}
\usepackage{changepage}

\usepackage{tensor}
\usepackage[colorinlistoftodos]{todonotes}
\usepackage{epsfig}
\usepackage{graphicx,color}
\usepackage{cancel}
\usepackage[utf8]{inputenc}
\usepackage{simpler-wick}
\usepackage{comment}
\usepackage{caption}
\usepackage{subcaption}
\usepackage{bm}

\newcommand{\veir}{\varepsilon_{\rm {IR}}}
\newcommand{\veuv}{\varepsilon_{\rm {UV}}}

\newcommand{\be}{\begin{equation}}
\newcommand{\ee}{\end{equation}}
\newcommand{\bea}{\begin{eqnarray}}
\newcommand{\eea}{\end{eqnarray}}
\newcommand{\balign}{\begin{align}}
\newcommand{\ealign}{\end{align}}

\newcommand{\bra}[1]{\left< #1 \right |}
\newcommand{\ket}[1]{\left | #1 \right >}
\newcommand{\braket}[1]{\left< #1 \right >}
\newcommand{\sandwich}[3]{\left< #1 \right | #2 \left | #3 \right >}

\newcommand{\bg}{\begin{gather}}
\newcommand{\foma}{\end{gather}}

\newcommand{\noopsort}[1]{}

\def\ve{\varepsilon}

\def\<{\langle}
\def\>{\rangle}

\def\d{\delta}

\def\({\left(}
\def\[{\left[}
\def\){\right)}
\def\]{\right]}

\def\sin{\hbox{sin}}
\def\ln{\hbox{ln}}

\def\Slash#1{{#1\!\!\!\slash}}

\newcommand{\ben}{\begin{eqnarray}}
\newcommand{\een}{\end{eqnarray}}

\newcommand{\bef}{\begin{figure}[htb]\centering}
\newcommand{\eef}{\end{figure}}

\usepackage[normalem]{ulem} 
\newcommand{\NOTE}[1]{\marginpar{\footnotesize\textbf{\color[rgb]{0.9,0,0.9}**NOTE**}}{\color[rgb]{0.9,0,0.9}\sf [#1]}}

\title{Modeling the TMD shape function in $J/\psi$ electroproduction}
\author[a,b]{Miguel G. Echevarria,}
\author[a,b,c]{Raj Kishore}
\author[a,b]{and Samuel F. Romera}
\affiliation[a]{Department of Physics, University of the Basque Country UPV/EHU, 48080 Bilbao, Spain}
\affiliation[b]{EHU Quantum Center, University of the Basque Country UPV/EHU}
\affiliation[c]{Center for Frontiers in Nuclear Science, Stony Brook University, Stony Brook, NY 11794-3800, USA}

\emailAdd{miguel.garciae@ehu.eus}
\emailAdd{raj.kishore@ehu.eus}
\emailAdd{samuel.fernandez@ehu.eus}

\date{\today}

\abstract{
The next-to-leading order hard function for quarkonium electroproduction is calculated within the framework of transverse-momentum-dependent (TMD) factorization in the low-transverse-momentum regime.
The structure of the TMD shape function in quarkonium leptoproduction is analyzed through its operator-level definition.
Particular attention is given to the convolution of the unpolarized TMD gluon distribution with the TMD shape function, thereby illustrating the latter's phenomenological role.
Building on this framework, we provide predictions for the unpolarized differential cross-section of $J/\psi$ electroproduction at the future Electron-Ion Collider in the region of small transverse momentum.
}

\preprint{preprint number}
\begin{document}
\maketitle

\section{Introduction}

The bound states of a heavy quark and a heavy anti-quark, known as quarkonia, have gained significant importance in recent years.
Since the production and decay of quarkonia occur with a very marked hierarchy of scales such that $m_Q \gg \Lambda_{\text{QCD}}$, where $m_Q$ is the mass of the heavy quark and $\Lambda_{\text{QCD}}$ is the energy at which QCD is no longer perturbative, all processes occurring at this scale, such as the production of a heavy-quark pair, can be calculated with perturbative QCD.
On the other hand, the formation of the bound state is not calculable using perturbative QCD, meaning that processes involving quarkonia are ideal tools for the study of the interplay between perturbative and non-perturbative (NP) QCD.
Another important aspect—and one of the main motivations for studying quarkonium production—is the extraction of the transverse-momentum-dependent (TMD) gluon distributions within the proton. While quark distributions have been extensively studied and are known with high precision, our knowledge of gluon distributions remains limited. In the search for processes where production or decay channels dominated by gluons prevail over those involving quarks, quarkonium has emerged as a valuable tool.

Despite decades of progress, many physical aspects of quarkonium dynamics remain poorly understood; for comprehensive discussions we refer to recent reviews~\cite{Boer:2024ylx, Lansberg:2019adr, Andronic:2015wma}.
A central open question concerns the mechanism governing the production and decay of heavy-quark bound states.
Several NP models and effective field theories have been proposed, among which non-relativistic QCD (NRQCD)~\cite{Bodwin:1994jh} has emerged as the most extensively studied framework.
NRQCD, as an effective field theory of QCD, provides a systematic factorization of quarkonium cross-sections into two distinct components: short-distance coefficients and long-distance matrix elements (LDMEs)~\cite{Nayak:2005rt,Nayak:2005rw,Kang:2011zza,Kang:2014tta}.
The former encodes the perturbatively calculable production of a heavy-quark pair at short distance.
In contrast, the LDMEs describe the NP transition of the heavy-quark pair into a physical quarkonium state, characterized by its spin $S$, orbital angular momentum $L$, total angular momentum $J$, and color configuration $[c]$ (octet $[8]$ or singlet $[1]$), conventionally denoted as $n = \,^{2S+1}L_J^{[c]}$.
Since LDMEs encapsulate hadronization probabilities, they cannot be computed perturbatively and must instead be extracted from experimental data~\cite{Butenschoen:2011yh,Shao:2014yta,Bodwin:2015iua,Feng:2018ukp,Brambilla:2022ayc}.
Furthermore, not all heavy-quark configurations contribute equally to quarkonium formation, as the LDMEs scale with powers of the heavy-quark relative velocity v~\cite{Braaten:1996ix}, such that practical calculations are truncated at finite order in the v-expansion.
Nonetheless, theoretical uncertainties persist, most notably the apparent process dependence of the extracted LDMEs, which challenges their assumed universality within NRQCD.

While a detailed discussion of these issues lies beyond the scope of the present work, they nonetheless provide important context for the study of quarkonium production at low transverse momentum that we pursue here.
In this case, a modification of NRQCD becomes necessary, specifically the investigation of the role played by soft gluons within the theoretical framework.
To address this, two approaches have been developed: pNRQCD~\cite{Brambilla:1999xf, Brambilla:2004jw} and vNRQCD~\cite{Luke:1999kz,Rothstein:2018dzq}.
In this work we focus on vNRQCD, where two types of gluons, ultrasoft and soft, are explicitly present.
Within this theoretical framework, a power counting of the interactions among the relevant degrees of freedom is carried out, such that heavy quarks can only undergo off-shell fluctuations at the ultrasoft scale, implying that single soft-gluon emissions from a heavy-quark line do not occur at leading power.
Therefore, soft gluons are not treated as dynamical degrees of freedom, but only as mediators whose self-interactions are already encoded in the construction of the effective field theory.

Recent developments~\cite{Echevarria:2019ynx, Fleming:2019pzj, Echevarria:2024idp, Copeland:2025vop} have clarified that a consistent formulation of TMD factorization in quarkonium production requires incorporating the interplay between soft radiation and bound-state formation, the latter being encoded in a process-dependent TMD shape function (TMDShF) or in a recently-defined TMD soft transition function (TMDSTF).
The necessity of introducing TMDShF was also demonstrated through a matching analysis of the $J/\psi$ electroproduction cross-section between the intermediate- and low-transverse-momentum regimes~\cite{Boer:2020bbd, Boer:2023zit}. 
In that framework, an explicit expression for the NLO TMDShF was obtained; however, it does not coincide with the definition derived via the operator-level construction in Ref.~\cite{Echevarria:2024idp}.
The origin of this discrepancy can be traced to a different choice of hard scale, a point that will be revisited in the current text.
Regarding the TMDSTF, while this work was being finalized, Ref.~\cite{Copeland:2025vop} appeared, introducing the so-called TMDSTF.
This object encodes the transition of color-octet (CO) heavy-quark pairs into color-singlet configurations through the emission of soft gluons.
In the power counting of vNRQCD, the operator defining the TMDSTF is expected to be leading with respect to the TMDShFs, since the color-singlet LDME scales as v$^3$, whereas the color-octet LDMEs scale as v$^7$.
At the same time, the presence of the gluon field renders the operator subleading in the strong coupling.
However, in this case, the strong coupling must be evaluated at the soft scale which is typically larger than its value at the hard scattering scale.
These considerations highlight the need for a more systematic investigation of the interplay between the relevant power-counting parameters, as well as of the possible dominance of TMDSTF contributions over those from TMDShFs, or viceversa, in the cross-section.
At present, the TMDSTF has only been computed perturbatively at leading order and without including TMD evolution, preventing any firm conclusions.
Our preliminary expectation, based on the numerical results shown in Ref.~\cite{Copeland:2025vop}, is that the TMDSTF contribution to the cross-section is strongly suppressed relative to the $^1S_0^{[8]}$ TMDShF contribution, owing to its behavior at small values of $b_T$. Nevertheless, as emphasized in Ref.~\cite{Copeland:2025vop}, this behavior is expected to be significantly modified once TMD evolution is consistently implemented.

The central aim of the present work is to investigate the TMD shape function contribution to the unpolarized $J/\psi$ electroproduction cross-section at low transverse momentum at the future Electron-Ion Collider~\cite{AbdulKhalek:2021gbh}.
To this end, we adopt the TMD factorization framework developed in Ref.~~\cite{Echevarria:2024idp}, in which the cross-section is expressed as the convolution of a TMD gluon distribution with the operator-defined TMDShF. Since the next-to-leading order (NLO) hard function has not previously been computed within this framework for the process under study, we also perform this calculation.
Concretely, we analyze electroproduction in the small-transverse-momentum region at lowest perturbative order, $\mathcal{O}(\alpha_{em}\alpha_s)$.
This corresponds to the color-octet states $^1S_0^{[8]}$, $^3S_1^{[8]}$, and $^3P_J^{[8]}$ (see, e.g., Ref.~\cite{Schuler:1997is}), collectively denoted as the $\mathcal{O}(\alpha_s)$ CO channel. Moreover, at $\mathcal{O}(\alpha_s)$, the contribution from $^3S_1^{[8]}$ state vanishes due to the Lorentz and spin structure of the two-vector initial state, so this channel only appears at higher order with an additional gluon in the final state.

The remainder of this paper is organized as follows. In Section~\ref{sec:csTMD}, we introduce the theoretical framework, review the factorized expression for the $J/\psi$ leptoproduction cross-section, and present the calculation of one-loop virtual QCD corrections in the small-transverse-momentum region.
As a result, we obtain the next-to-leading order (NLO) hard function.
Section~\ref{sec:TMDShF} is devoted to the modeling of the TMDShF, where we analyze its perturbative and non-perturbative behavior and investigate its role in the convolution with the gluon TMDPDF.
In Section~\ref{sec:theoretical-predictions}, we provide numerical predictions for the electroproduction cross-section in the kinematic regime of the Electron–Ion Collider.
Finally, our conclusions are summarized in Section~\ref{sec:conclusion}.


\section{Cross-section in TMD factorization} \label{sec:csTMD}

We present the TMD factorization of the unpolarized $J/\psi$ leptoproduction cross-section in terms of the hard function, the TMDShF and the gluon TMDPDF.
In this work, we neglect target-mass corrections, so the mass of the proton is considered as $M_{N} \approx 0$.

The semi-inclusive production of $J/\psi$ in a lepton-proton collision is defined as
\begin{eqnarray}\label{eq:process}
    \ell (k)+p(P)\to \ell (k')+J/\psi(P_{\psi})+X(P_X) \; ,
\end{eqnarray}
with the particle momenta indicated in parenthesis.
The differential cross-section of electroproduction, in the approximation of a single virtual photon exchange and in a photon frame, is given as
\begin{equation}
\begin{aligned}
\frac{d^5 \sigma}{d Q^2 d y d z d^2 \boldsymbol{P}_{\psi \perp}} & =\frac{\pi \alpha_{e m}^2 e_c^2}{2 Q^4} \frac{1}{S} \frac{1}{z} L_{\mu \nu}^{e} W^{\mu \nu}
\end{aligned}
\end{equation}
Here the azimuthal angle of the outgoing lepton has been integrated out, $e_c$ is the fractional electric charge of the charm quark and $Q^2 = - q^2$ with $q$ the virtual photon momentum such that $q = k - k'$. Regarding the parameters $y$ and $z$, along with the Bjorken variable $x_B$, they constitute the usual SIDIS invariants:
\begin{equation}
    x_B = \frac{Q^2}{2 P \cdot q} \; , \qquad y = \frac{P \cdot q}{P \cdot l} \; , \qquad z = \frac{P \cdot P_\psi}{P \cdot q} \; .
\end{equation}
Moreover, the tensor $L_{\mu \nu}$ denotes the leptonic tensor, given by
\begin{equation}
\begin{aligned} \label{eq:lept-tensors}
    L_{\mu \nu} & = 2\left(k_\mu k_\nu^{\prime}+k_\nu k_\mu^{\prime}-g_{\mu \nu} k \cdot k^{\prime}\right)+2 i \lambda_l \epsilon_{\mu \nu \rho \sigma} l^\rho q^\sigma \; ,
\end{aligned}
\end{equation}
where we have summed over the spin of the final lepton, and $\lambda_\ell = \pm 1$ is twice the helicity of the incoming lepton. Additionally, the contribution from the anti-symmetric (parity-violating) term will vanish for the unpolarized cross section.

The hadron tensor, $W^{\mu \nu}$, is defined as
\begin{equation}
    W^{\mu \nu} = \sum_{\,\,X}\!\!\!\!\!\!\!\!\int
    \int \frac{d^4b}{(2\pi)^4} e^{i q \cdot b}
    \sandwich{N}{J^{ \mu\dagger}(b)}{J/\psi,X}
    \sandwich{J/\psi,X}{J^\nu(0)}{N} \; ,
\end{equation}
where $J^\mu$ is the electromagnetic current in QCD. After the TMD factorization procedure, which is detailed in Ref.~\cite{Echevarria:2024idp}, we obtain that the hadron tensor, in transverse-momentum space, can be factorized in terms of the [n] hard function $H_{[n]}$,\footnote{In the current work, we denote the configuration of the heavy-quark pair as $[n]$ instead of $n$, avoiding confusion with the $n$-light-collinear direction.} 
the gluon TMDPDF $G_{g/N}$ and the $[n]$-TMDShF $S_{[n] \to J/\psi}$:

\begin{equation} 
\begin{aligned} \label{eq:Hadronic-Tensor}
W^{\mu\nu} 
& =\frac{1}{x y S} \,\delta(1-z)
\sum_{[n]} H_{[n]} \Gamma_{[n]}^{\dagger \mu \alpha'} \Gamma_{[n]}^{\nu\alpha}
\\
 \times&  
\int\mathrm{d}^{2}\bm{k}_{n\perp}\,\mathrm{d}^{2} \bm{k}_{s\perp}\,
\d^{2}(\bm{q}_\perp + \bm{k}_{n\perp} - \bm{k}_{s\perp})\,
G_{g/N , \alpha\alpha'}(x,\bm{k}_{n\perp};\mu,\zeta_A) \, 
S_{[n] \to J/\psi}(\bm{k}_{s\perp};\mu,\zeta_B)
\, ,
\end{aligned}
\end{equation}
where $\mu$ is renormalization scale, and $\zeta$'s are the rapidity scales, such that $\zeta_A \zeta_B = \mu_H^2$, with $\mu_H$ represents the hard factorization scale.
The tensor $\Gamma^{\nu \alpha}_{[n]}$ refers to the tensorial structure of the partonic process through the heavy-quark pair is produced; as can be seen, it depends on the state in which the pair is formed and is common to all orders in $\alpha_s$.

\subsection{NLO hard scattering for electroproduction}

Since the hard function encodes the short-distance dynamics—captured exclusively by the virtual corrections—and given that the tensor $\Gamma$ is the same at all orders of $\alpha_s$, the hard function can be expressed as $H_{[n]} = |C_{[n]}|^2$, obtained through a matching between the virtual part of the QCD cross-section for partonic heavy-quark pair production and its counterpart in the effective field theory.

As no explicit next-to-leading order (NLO) calculation of the virtual contribution to the electroproduction cross-section is available in the literature, we have performed this calculation ourselves and present the results here.
Our approach follows the procedure of Ref.~\cite{Maltoni:1997pt}, where the corresponding process was evaluated for photoproduction.
In our case, the introduction of an additional scale, $Q^2$, renders the calculation more intricate; for this reason, the details are provided in Appendix~\ref{sec:appendix-2}.

The final result is presented in the following way:
\begin{equation}
    \begin{aligned}
    \label{eq:virtual_corrections}
        \sigma_{[n]} & = \sigma_{\text{Born}[n]}
        \frac{\alpha_s}{2\pi}
        \left\{  
        \frac{\beta_0}{2\veuv} - C_A \left[ \frac{1}{\veir^2} + \frac{1}{\veir} \ln \left(\frac{\mu^2}{\mu_H^2} \right)\right]  -  \frac{\beta_0 + 2C_A}{2\veir}  + \mathcal D_{[n]}
        \right\} \; ,
    \end{aligned}
\end{equation}
where $\sigma_{\text{Born} [n]}$ is the tree level or Born cross-section (see Appendix \ref{sec:appendix-1}), $\beta_0 = 11 C_A/3-4T_F n_f/3$, and
\begin{equation}
\label{eq:muh}
    \mu_H^2 = \frac{(M^2 + Q^2)^2}{M^2} \; ,
\end{equation}
is a combination of the two scales involved in the hard scattering of leptoproduction.
This scale is not an ad-hoc choice, but it emerges naturally from the perturbative calculation.\footnote{
We note that the hard scale of the process is clearly then $(M^2 + Q^2)/M$.
This scale does not coincide with the hard scale adopted in previous studies of quarkonium electroproduction, where the virtual corrections to the cross-section at NLO were not computed explicitly.
In particular, Ref.~\cite{Boer:2023zit} illustrates this ambiguity: there, the TMD shape function was extracted via a matching procedure between the low- and intermediate-transverse-momentum regimes, yielding a distribution that differs from the operator-level definition of the TMDShF given in Ref.~\cite{Echevarria:2024idp}.
The origin of this discrepancy lies in the hard scale: Ref.~\cite{Boer:2023zit} employs the ansatz $\sqrt{M^2+Q^2}$, instead of the expression obtained in this work in Eq.~(\ref{eq:Finite-term}), leading to a TMDShF with an unphysical $Q^2$ dependence.
This in turn motivated a discussion regarding how to remove such dependence, noting that the TMDShF is intrinsically a soft distribution.
We have explicitly verified that adopting the hard scale $\mu$ as in \eqref{eq:muh} in the analysis of Ref.~\cite{Boer:2023zit} eliminates the spurious $Q^2$ dependence, yielding a TMDShF that agrees with the operator-level result of Ref.~\cite{Echevarria:2024idp}.
}
Therefore, the renormalization scale in the hard factor, $\mu$, is chosen to be proportional to $\mu_H$ to minimize logarithms in the perturbative calculation.
Here the strong coupling is bare, that is, the UV divergence disappears after renormalization.
Since the IR divergences must match in both the effective theory and QCD, it can be easily verified that these divergences arise purely from the contribution of the gluon TMDPDF and the TMDShF at NLO.
The remainder will be a finite term that contains no divergences and constitutes the hard function:
\begin{equation}
    \begin{aligned}
        H_{[n]} & = 1 + \frac{\alpha_s(\mu)}{2 \pi} \,
         H^{(1)}_{[n]} \; ,
    \end{aligned}
\end{equation}
where the most general form of the one-loop term is the following
\begin{equation}
\begin{aligned}
\label{eq:Finite-term}
     H^{(1)}_{[n]} & = - \frac{C_A}{2} \ln ^2\left(\frac{\mu^2}{\mu_H^2}\right)-C_A \ln \left(\frac{\mu^2}{\mu_H^2}\right) + f_{1 [n]}  \ln(2) + f_{2[n]} \pi^2 + f_{3[n]} \\
     & + f_{4 [n]} \ln(\rho + 1 ) + f_{5 [n]} \tanh^{-1} \left( \sqrt{\frac{\rho}{\rho+1}} \right) + f_{10 [n]} \sin^{-1} \left( \sqrt{\rho} \right)+ f_{6 [n]} \text{Li}_2 (-1 - 2 \rho) \\
     & + f_{7[n]} \left[ \text{Li}_2 (-2 \sqrt{\rho(\rho+1)} - 2 \rho) + \text{Li}_2 (2 \sqrt{\rho(\rho+1)} - 2 \rho) \right] + f_{8[n]} I_1 + f_{9[n]} I_2 \; .
\end{aligned}
\end{equation}
Here the coefficients $f_{[n]}$ are given by rational functions of $\rho \equiv Q^2/M^2$ and depend on the Casimir operators $C_A$ and $C_F$, while $I_1$ and $I_2$ denote one-loop integrals that we have evaluated numerically; their explicit forms are provided in Appendix~\ref{sec:appendix-2}. The corresponding expression for $H^{(1)}_{[n]}$ in photoproduction is readily obtained by taking the limit $Q \to 0$ (i.e., $\rho \to 0$) of the above result, which reproduces the known result in the literature, see e.g. Ref.~\cite{Maltoni:1997pt}.\footnote{
One should notice that the hard function corresponding to the $^3P_1^{[8]}$ state develops a singularity of the form $1/(M^2 - Q^2)$ at $M^2 = Q^2$.
Since our analysis focuses on the kinematic region where $z \to 1$, the condition $Q^2 > M^2$ appears physically sensible, as for $Q^2 < M^2$ the available energy would be insufficient to produce the final state.
Given that our results for the $^3P_0^{[8]}$ and $^3P_2^{[8]}$ channels are consistent with the literature in the photoproduction limit, we restrict our cross-section analysis and subsequent EIC predictions to the region $Q^2 \gtrsim M^2$.
}

\section{Analysis of the TMDShF}\label{sec:TMDShF}

In this section, we present the theoretical framework and notation relevant to the TMD shape function within the context of TMD evolution.
These definitions form the basis for the numerical analyses discussed in the subsequent section.
In particular, we specify the model employed for the TMDShF and illustrate its implementation in the convolution with the gluon TMDPDF, which constitutes the principal theoretical ingredient in the computation of the quarkonium leptoproduction cross section.

\subsection{Evolution and non-perturbative behavior}

As discussed in the previous section, the QCD dynamics of the process are encoded in three independent functions: the unpolarized gluon TMDPDF, the $[n]$-TMDShF, and the Collins–Soper kernel $\mathcal{D}$.
Each of these functions is modeled separately.
At small values of $b_T$, the distributions are computed perturbatively in terms of $\alpha_s$ and collinear PDFs, while at larger $b_T$ they are described by parametrized functions with a few free parameters determined from fits to experimental data.
We now provide a detailed account of the modeling of the TMDShF.

The $[n]$-TMDShF describes the effect on the quarkonium hadronization of the (u)soft-gluon exchange between the heavy quarks in the $[n]$ bound state and the (u)soft modes of the process.
It is defined at small-$b_T$ by the following expression
\begin{equation}
\label{eq:OPE}
    S_{[n]}(b_T;\mu,\zeta) = \sum_{[m]} C_{[m]}^{[n]}(b_T;\mu,\zeta) \frac{\braket{\mathcal{O}^{[m]}}(\mu)}{N_{pol}^{(J)}}\; .
\end{equation}
Here $\braket{\mathcal{O}^{[m]}}$ denotes the $[m]$-LDME, $C_{[m]}^{[n]}$ is the matching coefficient relating the $[n]$-TMDShF to the $[m]$-LDME (see Appendix~\ref{sec:appendix-1}), and $N_{\text{pol}}^{(J)} = 2J+1$ is the number of polarizations of the state $[m]$ with total angular momentum $J$.
The renormalization scale $\mu$ and the rapidity scale $\zeta$ are defined at the point where the non-perturbative input for the TMDShF is specified.

In practical applications, one has to evaluate these objects at the scale where the hard collision takes place. Therefore, the evolution from the initial point $(\mu_i,\zeta_{i})$ of the equation (\ref{eq:OPE}) to a final point $(\mu_f,\zeta_{f})$ is necessary. This can be obtained by solving the pair of renormalization group equations and the solution gives the following general expression for the TMDShF at $(\mu_f,\zeta_{f})$
\begin{equation}
\begin{gathered} \label{eq:OPE2}
    S_{[n]} ( b_T; \mu_f, \zeta_{f} )  = S_{[n]} (b_T; \mu_i, \zeta_{i}) \, R(b_T; (\mu_i, \zeta_{i}) \to (\mu_f, \zeta_{f})) \; , \\
     \text{with} \quad R = \exp \left[  \int_{\mu_i}^{\mu_f} \frac{d \bar{\mu}}{\bar{\mu}} \, \gamma_S \left( \bar{\mu}, \zeta_{f} \right) - \mathcal D \left( b_T;\mu_i \right) \,\ln\frac{\zeta_{f}}{\zeta_{i}} \right]  \; .
\end{gathered}
\end{equation}
Within the TMD evolution factor $R$, $\gamma_S$ is called the TMDShF anomalous dimension and $\mathcal D$ is called also the rapidity anomalous dimension.
The latter describes the soft-gluon exchange between hadrons and can be determined from the rapidity divergent part of the soft function at small values of $b_T$~\cite{Echevarria:2015byo}.
In fact, its perturbative behavior, named $\mathcal D_{pert}$, is currently known at N$^3$LO~\cite{Duhr:2022yyp}.
We emphasize that LDMEs must evolve from the scale at which they are extracted to the scale $\mu_i$, as we discuss below.
Moreover, since the anomalous dimensions of the two scales are correlated, the mutual dependence is explicitly worked out through the so-called cusp anomalous dimension $\Gamma_{cusp}$, such as the TMDShF anomalous dimension can be written as
\begin{equation}
   - \gamma_S(\mu,\zeta) =  \Gamma_{cusp}(\mu) \, \ln \, \zeta + \gamma_s(\mu) \; ,
\end{equation}
where $\Gamma_{cusp}$ is currently known at N$^4$LO~\cite{Herzog:2018kwj}. The anomalous dimension $\gamma_s$ refers to the finite part of the renormalization of the shape function and is known at NLO~\cite{Echevarria:2024idp}.
We direct the reader to the Appendix \ref{sec:appendix-1}, where there is a summary of all these expressions in the truncated perturbation theory.

In Eq.~(\ref{eq:OPE2}), the initial and final scales must be specified. For the final scales, we adopt the hard factorization scale of the process, $\mu_H$.
Consistency of the TMD factorization framework requires that the rapidity scales satisfy $\zeta_A \zeta_B = \mu_H^2$, which we enforce by setting $\zeta_A = \mu_H^2$ and $\zeta_B = 1$, so that $(\mu_f, \zeta_f) = (\mu_H, 1)$.
For the initial scales, the suppression of large logarithmic contributions motivates the choice $\mu_i \sim 1/b_T$ and $\zeta_i = 1$.
Although $\zeta_B$ is dimensionless and the shape function does not generate large rapidity logarithms, it remains relevant to assess the impact of rapidity evolution in scale-variation studies. In particular, any shift in the initial rapidity scale of the TMDPDF entails a corresponding adjustment in the initial $\zeta$ of the TMDShF, such that the constraint $\zeta_A \zeta_B = \mu_H^2$ is maintained.

To separate perturbative and NP contributions, one defines small and large $b_T$ through $b_T^*(b_T)$ that freezes above $b_{T\text{max}}$ and equals to $b_T$ for small $b_T$:
\begin{equation}\label{eq:btstar}
    b_T^* (b_T) = \frac{b_c(b_T)}{\sqrt{1 + \frac{b_c^2(b_T)}{b_{T\text{max}}^2}}}  \qquad \text{with} \qquad b_c(b_T) = \sqrt{b_T^2 + \frac{b_0^2}{\mu_H^2}} \; ,
\end{equation}
where $b_c$ fixes the lower limit $b_{T\text{min}} \equiv b_0/\mu_H \leq b_T$ which marks the point where $\mu_b \equiv b_0/b_T$ becomes larger than $\mu_H$.
Therefore, this $b_T$-prescription ensures that $b_{T\text{min}} \leq b_T \leq b_{T\text{max}}$.
For $b_0$, see Appendix \ref{sec:appendix-1}.
It is easy to check that $b_T^*$ approaches $b_{T\text{min}}$ and $b_{T\text{max}}$ slowly, and there is not a clean separation of both perturbative and NP regions, so this can affect the functions of $b_T^*$ as the shape function, such that the NP distribution contributes to the perturbative one.
However, since the point where perturbation theory starts to fail is defined in the range of $0.5-1.5$ GeV$^{-1}$, but not a fixed value, we will not dwell on it at this point and we will use the expressions of Eq.~(\ref{eq:btstar}).

According to the above discussion, the TMDShF in the $b_T^*$-prescription at the point of the hard factorization is as follows 
\begin{equation} \label{eq:shape_function}
\begin{aligned}
    S_{[n]}(b_T;\mu_H,1) & = \sum_{[m]} C_{[m]}^{[n]}(b_T^*;\mu_{b^*},\zeta_{B,b}) \frac{\braket{\mathcal{O}^{[m]}}(\mu_{b^*})}{N_{pol}^{(J)}} \times e^{-S_{\text{NP}}(b_T)} \\
    & \times \exp\left[ - \int^{\mu_H}_{\mu_{b^*}} \frac{d \bar{\mu}}{\bar{\mu}} \gamma_s(\bar{\mu}) + \mathcal{D}_{pert}(b_T^*;\mu_{b^*}) \ln \, \zeta_{B,b}  \right] \; ,
\end{aligned}
\end{equation}
with
\begin{equation}
    S_{\text{NP}}(b_T) = \mathcal{S}_{\text{NP}}(b_T) + \mathcal D_{\text{NP}}(b_T) \; ,
\end{equation}
where $\mu_{b^*} = b_0/b_T^*$ and $\zeta_{B,b} = \zeta_{B,i}$.
The functions $\mathcal{S}_{\text{NP}}$ and $\mathcal{D}_{\text{NP}}$ encode the NP contributions to the TMDShF and the Collins–Soper kernel, respectively.
Although several parameterizations have been proposed in the literature, primarily in the context of TMDPDFs, these typically adopt a Gaussian functional dependence on $b_T$.
Since our focus is not on exploring the interplay between perturbative and non-perturbative physics, but rather on quantifying the impact of the TMD shape function, we adopt this widely used Gaussian form:
\begin{equation}
    \begin{aligned}
        \mathcal S_{\text{NP}}(b_T) =   B_S \,b_T^2 \; \quad \text{and} \quad 
        \mathcal D_{\text{NP}} (b_T) =   A_S \, \ln \left(\frac{1}{\zeta_{B,0}} \right)\,b_T^2 \; ,
    \end{aligned}
\end{equation}
where the parameters $B_S$, $A_S$ (which have dimensions of GeV$^2$)
and $\zeta_{B,0}$ (dimensionless) have never been extracted from the experimental data, and thus can only guess their values, giving predictions by varying them in a reliable range.
The parameter $\zeta_{B,0}$ determines the separation between the small-$b_T$ and large-$b_T$ regions of the TMD shape function. In the following, we define the perturbative Sudakov factor for the TMDShF based on Eq.~(\ref{eq:shape_function}):
\begin{equation}
    S_A(b_T; \mu_H,1) = \int_{\mu_{b^*}}^{\mu_H} \frac{d\bar{\mu}}{\bar{\mu}} \gamma_s(\bar{\mu}) - \mathcal{D}_{pert}(b_T^*,\mu_{b^*}) \ln \, \zeta_{B,b} \; .
\end{equation}

\subsection*{Evolution of the LDMEs}

\begin{figure}[t]
    \centering
    \includegraphics[width=\textwidth]{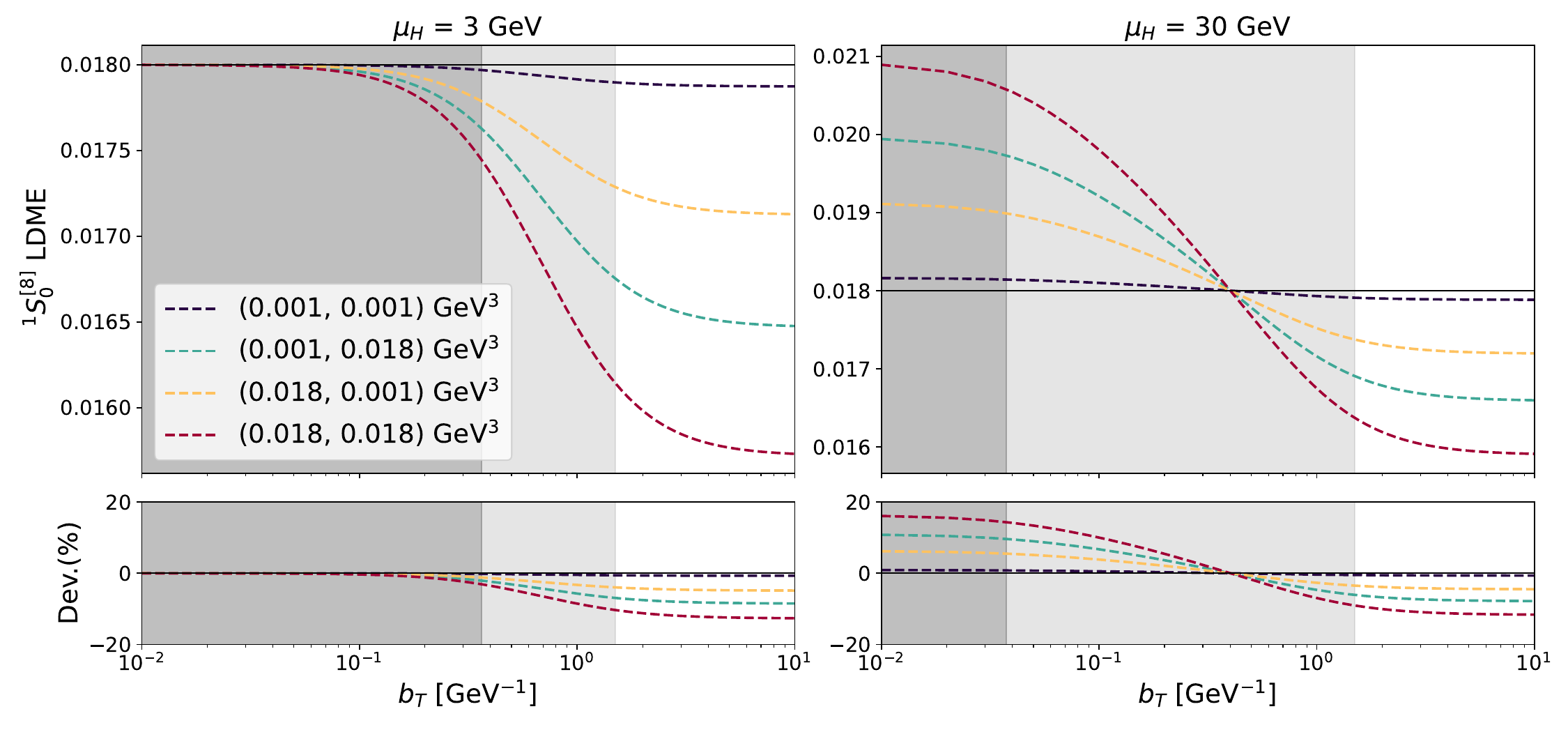}
    \caption{
    Evolution of the $\, ^1S_0^{[8]}$ LDME as a function of $b_T$.
    The middle band shows the perturbative region ($b_{T\text{min}} \leq b_T \leq b_{T\text{max}}$), and the horizontal black line corresponds to the value of the extracted $\, ^1S_0^{[8]}$ LDME.
    The dashed lines represent the evolved LDME considering the pair of values $\left( \scriptstyle \left\langle \, ^1P_1^{[1]} \, \right\rangle , \left\langle \, ^1P_1^{[8]} \, \right\rangle \right)$.}
    \label{fig:LDME_evolution}
\end{figure}

In Eq.~(\ref{eq:shape_function}), the LDMEs are evaluated at the scale $\mu_{b^*}$, whereas they are typically obtained at the quarkonium transverse-mass scale, $m_T = \sqrt{M^2 + P_{T\psi}^2}$, with $P_{T\psi}$ denoting the quarkonium transverse momentum.
Consequently, the LDMEs must be evolved from $m_T$ to $\mu_{b^*}$ using the evolution equations summarized in Appendix~\ref{sec:appendix-1}.
At tree level, the evolution is trivial, corresponding to the LDME evaluated directly at the final scale.
At one loop, however, the evolution is off-diagonal, meaning that the running of a given LDME is influenced predominantly by the LDMEs of other states.
In particular, the evolution of the $\, ^1S_0^{[8]}$ LDME is driven by the LDMEs of the $\, ^1P_1^{[1]}$ and $\, ^1P_1^{[8]}$ states.
Since experimental determinations of these LDMEs are not available, we vary them within a reasonable range to assess the impact of their contribution to the evolution of the $\,^1S_0^{[8]}$ LDME.

In Fig.~\ref{fig:LDME_evolution}, we display the evolution of the $\, ^1S_0^{[8]}$ LDME as a function of $b_T$, where the evolution scale is chosen as $\mu = \mu_{b^*}$.
The panels correspond to two representative hard scales, and the colored curved denote different input values do the P-states LDMEs entering the evolution equation of the $\, ^1S_0^{[8]}$ state.
For this analysis, we consider the SV set~\cite{Sharma_2013}, where $\braket{^1S_0^{[8]}}$ is equal to 0.018 GeV$^3$; although this value has been extracted at the scale $m_T$, here we consider the approximation $m_T \approx M$.
We assume that the P-state LDMEs are suppressed with respect to the S-state LDMEs, and we also analyze the extreme case in which they are equal.
The strength of this suppression is controlled by the choice of the P-wave matrix elements: small values produce only mild deviations, while larger values result in a more pronounced decrease.
This sensitivity illustrates the interplay between S-wave and P-wave channels in the evolution kernel.
The comparison between two panels highlights the scale dependence.
The lower plots quantify the deviation with respect to the extracted $\, ^1S_0^{[8]}$ LDME. 
In the intermediate and NP domains, the deviation grows, reaching up to $\mathcal{O}(10 \% - 20\%)$, depending on the chosen P-wave inputs.
Furthermore, we found that for P-wave LDMEs exceeding $\sim 0.1~\text{GeV}^3$, the $^1S_0^{[8]}$ LDME becomes negative in the intermediate and large-$b_T$ regions.
Although it is clear we do not have any extraction for these LDMEs for $J/\psi$, after this naive analysis we conclude that the evolution is sensitive to the values of the P-states LDMEs and that this sensitivity is amplified at large $b_T$ and higher hard scales.
Therefore, for a precise phenomenological analysis, the error induced by treating the LDMEs as constants could be significant, but we do not include it in this work. As can be seen in Eq.~(\ref{eq:LDME_evolution}), the functional form of the evolution for the P-states is the same as for the S-states. Therefore, if we use the same range of values for the LDMEs of the D states used in this analysis, we expect the evolution to be  formally identical in structure for the P -states.

\subsection*{TMDShF within TMD evolution}
We emphasize that for $\zeta_{B,b} = 1$ there is no contribution from $\mathcal{D}$; in other words, the TMDShF exhibits no evolution in rapidity.
However, this is not strictly correct: from the perspective of the renormalization group evolution equations, the TMD shape function evolves from an initial to a final rapidity scale.
The apparent discrepancy arises from the choice of the final scale in the TMD factorization of the cross-section.
For instance, one could choose $\zeta_A = 2 \mu_H^2$ and $\zeta_B = 1/2$ such that the final rapidity scale does not coincide with the natural scale of the logarithms appearing in the TMDShF at NLO.
Accordingly, we do not fix $\zeta_{B,b}$, but instead study its impact along with the variation of the relevant scales by replacing $\mu_H \to C_H  \mu_H$ and $\mu_{b^*} \to C_{b^*}  \mu_{b^*}$, with $C_H, C_{b^*}, \zeta_{B,b} \in \{0.5, 1.0, 2.0\}$.
Moreover, from the expressions above, we observe that $S_A$ vanishes at leading-logarithm accuracy (see Table~\ref{tab:orders_accuracy}).
At NNLL$^\prime$ and NLL, $S_A$ is identical when $C_H = C_{b^*} = \zeta_{B,b} = 1$, so its contribution is unchanged when the perturbative order is increased, at this stage.
Differences appear only when the scales are varied, i.e., for $C_H, C_{b^*}, \zeta_{B,b} \neq 1$.

\begin{table}[t]
    \centering
    \begin{tabular}{|c|c|c|c||c|}
        \hline
         acc. & $\Gamma_{cusp}$ & $\gamma_s$ & $D_{pert}$ &  $C^{[n]}_{[m]}$ \\ \hline
        LL & $\alpha^{1}_{s}$ & $\alpha^{0}_{s}$ & $\alpha^{0}_{s}$ &  $\alpha_s^0$ \\ \hline
        NLL & $\alpha^{2}_{s}$ & $\alpha^{1}_{s}$ &  $\alpha^{1}_{s}$ &  $\alpha_s^0$ \\ \hline
        NNLL$'$ & $\alpha^{3}_{s}$ & $\alpha^{1}_{s}$ & $\alpha^{2}_{s}$ & $\alpha_s^1$  \\ \hline
    \end{tabular}
    \caption{Perturbative orders of accuracy for resummation in impact parameter space.}
    \label{tab:orders_accuracy}
\end{table}

\begin{figure}[t]
    \centering
    \begin{subfigure}[b]{0.49\textwidth}
        \includegraphics[width=\textwidth]{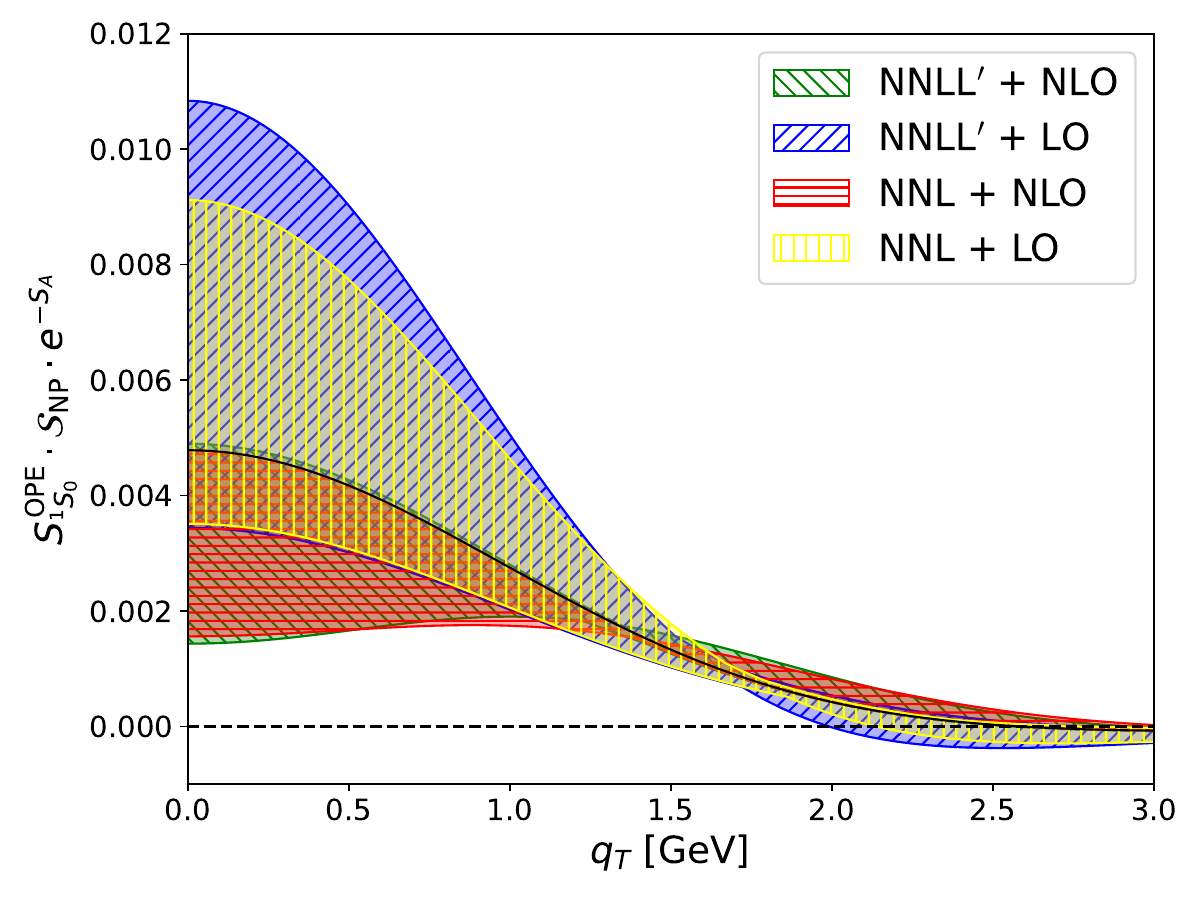}
    \end{subfigure}
    \hfill
    \begin{subfigure}[b]{0.49\textwidth}
        \includegraphics[width=\textwidth]{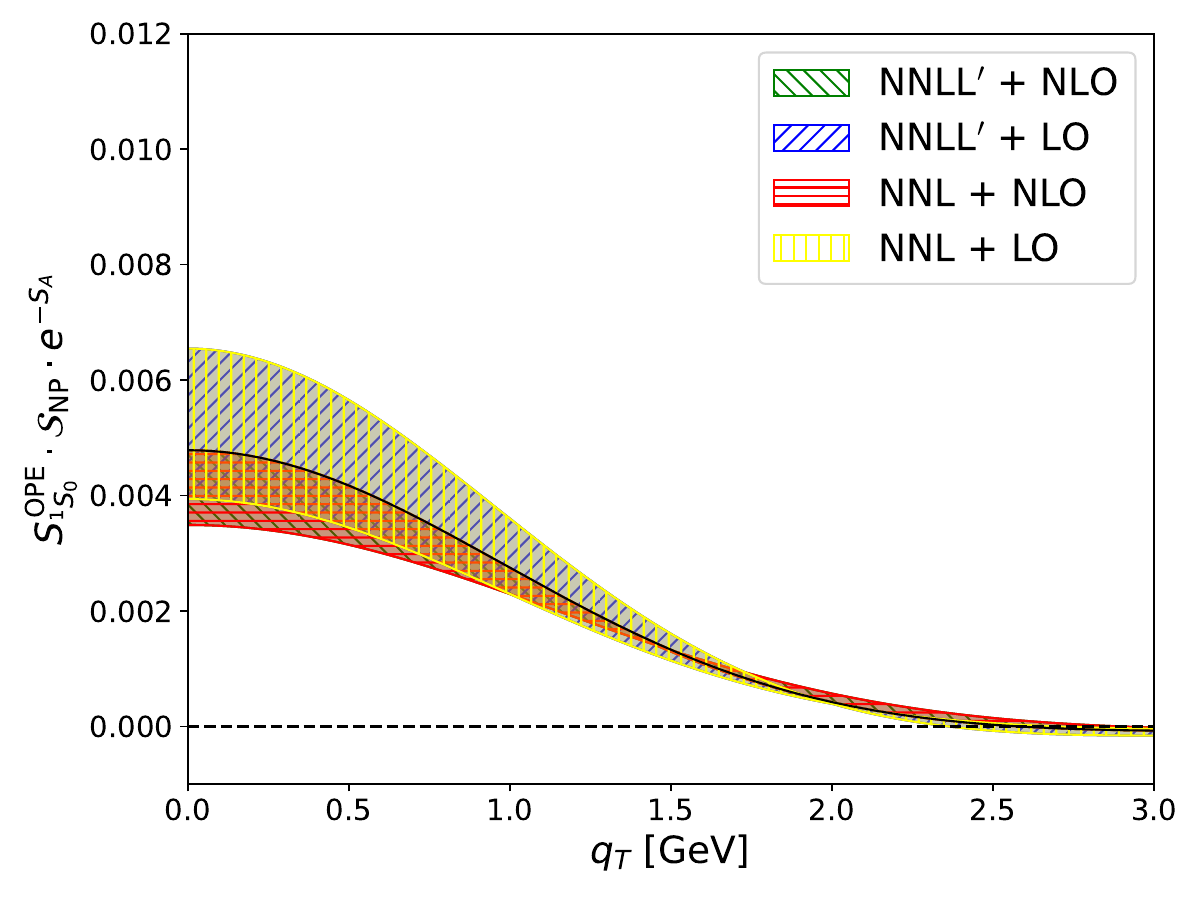}
    \end{subfigure}
    \caption{$q_T$-spectrum of the theoretical uncertainty of the $\, ^1S_0^{[8]}$ TMDShF for several levels of accuracy and for $\zeta_{B,b} \in \{2,1,0.5\}$ (left) and $\zeta_{B,b} = 1$ (right). Bands are obtained by varying the renormalization scale and the rapidity scale by a factor of 2. The black line represents the default value.}
    \label{fig:TMDShF-qt}
\end{figure}

In Fig.~\ref{fig:TMDShF-qt} we show the $\,^1S_0^{[8]}$ TMDShF in the $q_T$-space at $\mu_H = 15$ GeV, and $b_{T\text{max}} = 1.5$ GeV$^{-1}$ at the levels of accuracy illustrated in Tab.~\ref{tab:orders_accuracy}.
To be consistent with the factorization theorem, we calculate the theoretical curves up to $q_T = \mu_H/2$, but in the figure is shown only the scale variation up to $q_T = 3$ GeV.
For the NP parametrization, we fix $\zeta_{B,0} = 1$ (i.e., $\mathcal D_{\text{NP}} = 0$), and we fix $B_S = 0.5$ GeV$^2$.
The running of the strong coupling is implemented at N$^3$LO with $\Lambda_{\text{QCD}} = 0.239$ GeV$^{-1}$, while the flavor number scheme is the one fixing $n_f = 4$.
The bands come from varying both $C_{b^*}$ and $\zeta_{B,b}$ with a 9-point variation, and keeping the largest variation for each point in the $q_T$-spectrum. 
The bands at NNLL$'$ + NLO and at NNL + NLO are the narrower bands, as they include a higher perturbation order for $C_{[m]}^{[n]}$.
At a fixed resummation order in $S_A$, from comparing the green and blue bands or the red and yellow bands, we observe a reasonable reduction ($\sim 48\%$) in the theoretical uncertainty.
As we can see in the right picture, the most decreasing in the uncertainty is coming from varying only $\mu_{b^*}$, which directly affect the NLO order of $C_{[m]}^{[n]}$ through $\ln(\mu^2 b_T^2)$.
The difference with the left picture comes from the contribution of $\mathcal D_{pert}$ which enters in the game when $\zeta_{B,b} \neq 1$, and the thicker bands are obtained when $\zeta_{B,b} = 0.5$ as it flips the sign of the $\mathcal D_{pert}$ term in the $S_A$.
It is clear that the theoretical uncertainty gets reduced as we increase the resummation accuracy by including more perturbative ingredients, especially in $C_{[m]}^{[n]}$.

\begin{figure}[t]
    \centering
    \begin{subfigure}[b]{0.49\textwidth}
        \includegraphics[width=\linewidth]{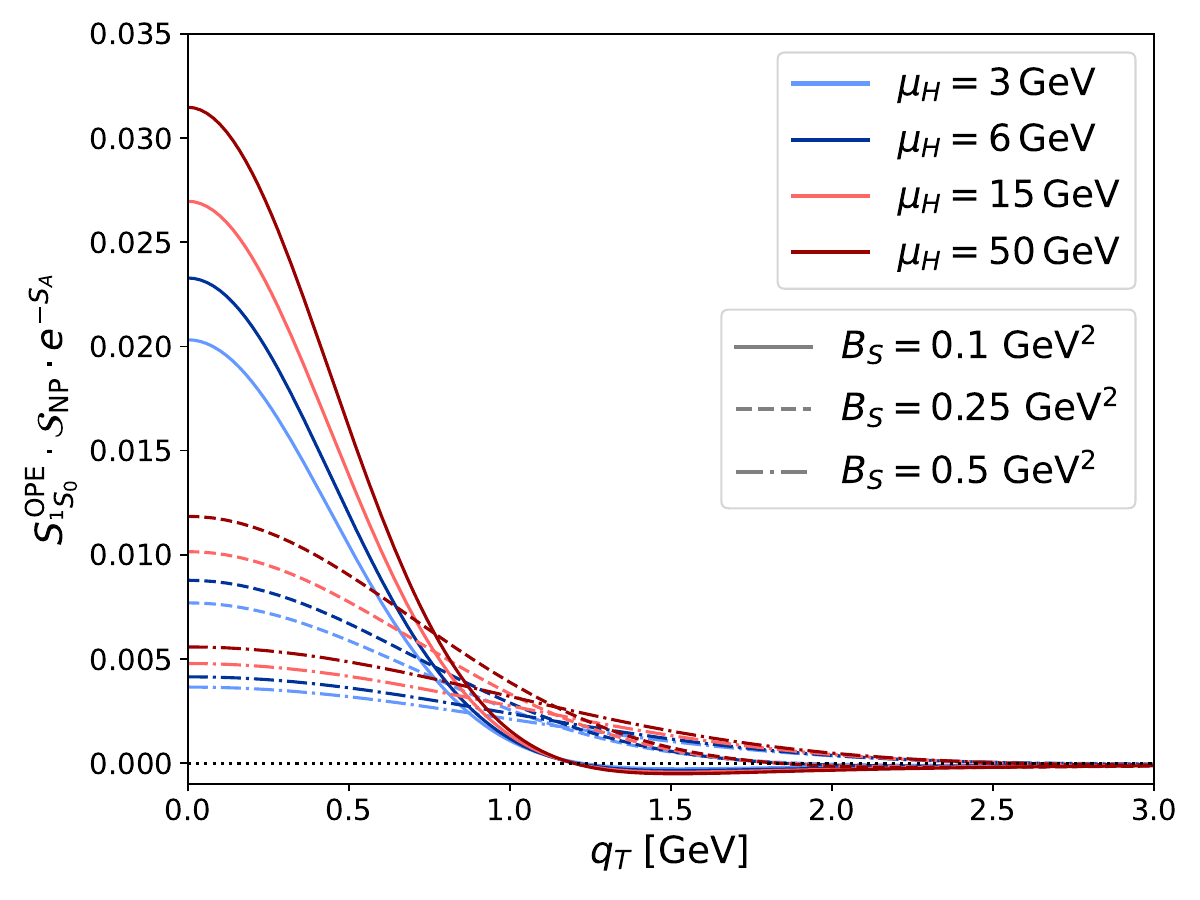}
    \end{subfigure}
    \hfill
    \begin{subfigure}[b]{0.49\textwidth}
        \includegraphics[width=\textwidth]{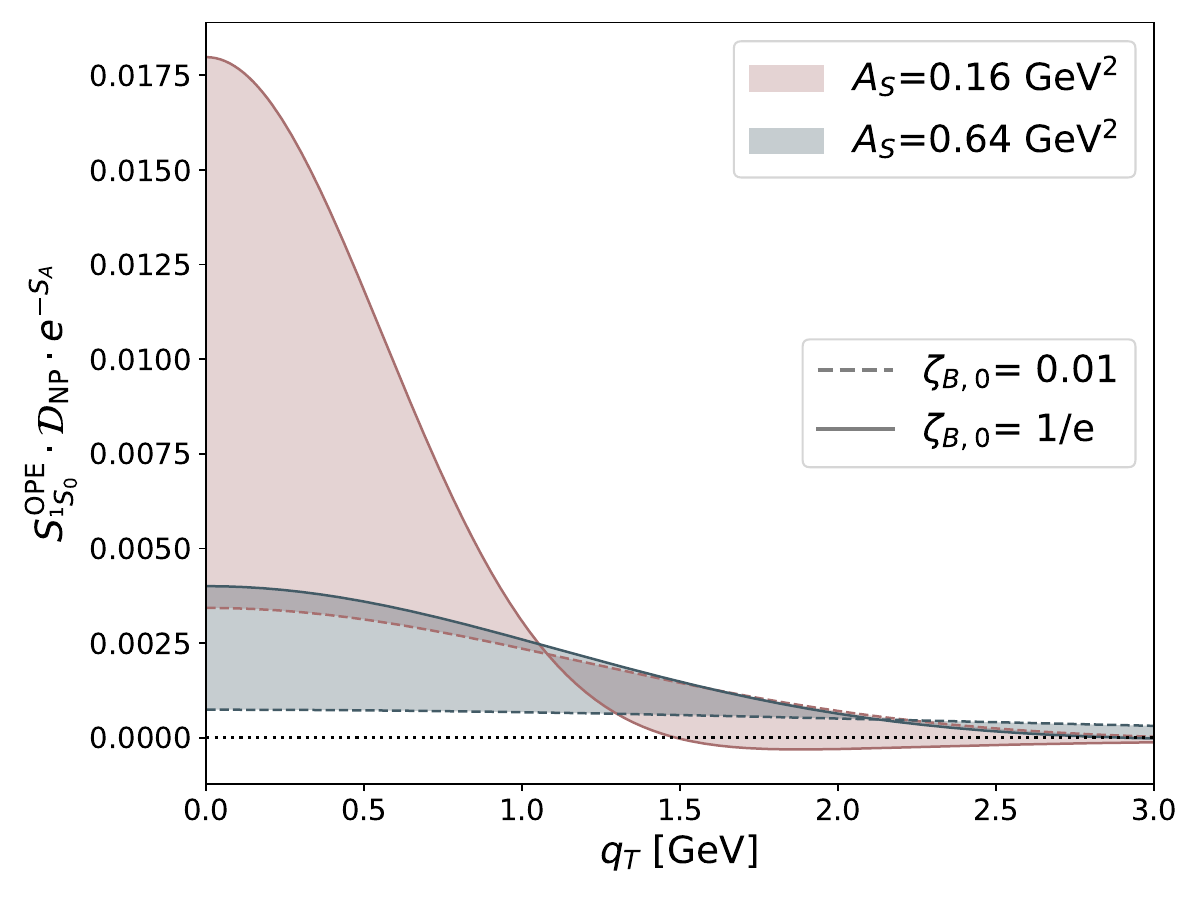}
    \end{subfigure}
    \caption{$q_T$-spectrum of the $\, ^1S_0^{[8]}$ TMDShF considering $S_{\text{NP}} = \mathcal S_{\text{NP}}$ (left) and $S_{\text{NP}} = \mathcal D_{\text{NP}}$ (right) for several values of the NP parameters.}
    \label{fig:TMDShF-energy_BS}
\end{figure}

In the left panel of Fig.~\ref{fig:TMDShF-energy_BS}, we display the TMDShF at NNLL$^\prime$+NLO accuracy for various energy scales and NP parametrizations, fixing $\zeta_{B,0}=1$.
The choice $\mu_H = M = 3$ GeV corresponds to the photoproduction hard scale, which will be discussed in the following section.
We observe that the peak of the distribution increases with the scale.
Furthermore, larger values of $B_S$ yield flatter distributions, with a suppressed magnitude at low $q_T$.
Conversely, for smaller values of $B_S$, the distribution exhibits a pronounced peak at low $q_T$ followed by oscillatory behavior at higher $q_T$.
This behavior originates from the fact that decreasing $B_S$ induces a flatter non-perturbative tail in $b_T$-space, which in turn generates oscillations after the Fourier transform.
In the right plot of Fig.~\ref{fig:TMDShF-energy_BS} we show the effect of $\mathcal D_{\text{NP}}$ on the TMDShF at NNLL$'$ + NLO for $\mu_H = 30$ GeV.
The parametrization of two values is involved here; $A_S$ and $\zeta_{B,0}$.
Since there is total arbitrariness in these parameters, we consider $A_S$ controlling the width of the Gaussian and using $\zeta_{B0}$ for the uncertainty of the NP parametrization.
Since $\zeta_{B,0}$ must lie within $(0,1]$ to ensure a decaying exponential behavior of $\mathcal D_{\text{NP}}$, we present results as uncertainty bands for $A_S={0.16,0.64}$~GeV$^2$, spanning the range $\zeta_{B,0} = 10^{-2}$ [$\ln(1/\zeta_{B,0}) = 4.61$] to $\zeta_{B,0} = 1/e$ [$\ln(1/\zeta_{B,0}) = 1$].
As expected, smaller values of $A_S$ correspond to larger uncertainty bands.

\subsection{Effect on the cross-section}

Having established the model for the TMD shape function, we now turn to its phenomenological application.
In this section, we analyze the convolution $\mathcal C[f_1^g S_{[n]}]$, which constitutes the most important contribution to the quarkonium leptoproduction cross-section in the TMD regime.
Before addressing the convolution itself, we first specify the model adopted for the gluon TMDPDF, $f_1^g$.

At the small-$b_T$ region ($b_T \ll 1/\Lambda_{\text{QCD}}$), the gluon unpolarized TMDPDF can be given by an operator product expansion in terms of collinear PDFs, $f_{a/N}$, and perturbatively calculable Wilson coefficients, $C_{g/a}$:
\begin{equation} \label{eq:tmd_ope}
    f_1^g(x, b_T, \mu, \zeta) = \sum_{a=q,\bar{q},g} C_{g/a}(x, b_T; \mu, \zeta) \otimes f_{a/N}(x, \mu) \; ,
\end{equation}
where we use the $C_{g/a}$ at NLO, see e.g., the Ref.~\cite{Echevarria:2015uaa}.
Moreover, the evolved TMDPDF at an arbitrary scales $(\mu_f, \zeta_f)$ is obtained from its value at the initial scales $(\mu_i, \zeta_i)$, using the evolution kernel:
\begin{equation} \label{eq:evolved_tmdpdf}
\begin{gathered}
    f_1^g(x, b_T; \mu_f, \zeta_f) = f_1^g(x, b_T; \mu_i, \zeta_i) \, R(b_T; (\mu_i, \zeta_{i}) \to (\mu_f, \zeta_{f})) \\
    \text{with} \quad R = \exp \left[  \int_{\mu_i}^{\mu_f} \frac{d \bar{\mu}}{\bar{\mu}} \, \gamma_F \left( \bar{\mu}, \zeta_{f} \right) - \mathcal D \left( b_T;\mu_i \right) \,\ln\frac{\zeta_{f}}{\zeta_{i}} \right]  \; ,
\end{gathered}
\end{equation}
where the gluon TMDPDF anomalous dimension is defined as
\begin{equation}
    -\gamma_F(\mu,\zeta) = \Gamma_{cusp}(\mu) \, \ln\left(\frac{\zeta}{\mu^2} \right) + \gamma_V(\mu) \; .
\end{equation}
The function $\gamma_V$ refers to the finite part of the renormalization of the vector form factor, which is known at N$^3$LO~\cite{Lee:2022nhh}.

According to the Eq.~(\ref{eq:evolved_tmdpdf}) and the $b_T^*$-prescription discussed above, we defined the gluon unpolarized TMDPDF at the point of the hard factorization $(\mu_H,\zeta_A = \mu_H^2)$ as follows
\begin{equation} \label{eq:gluon_tmdpdf}
\begin{aligned}
    f_1^g(x,b_T;\mu_H, \mu_H^2) & = \sum_{a=q,\bar{q},g} C_{g/a}(x, b_T^*; \mu_{b^*}, \mu_{b^*}^2) \otimes f_{a/A}(x, \mu_{b^*}) \times e^{-S_{\text{NP}}(x,b_T)} \times e^{-S_A} \; .
\end{aligned}
\end{equation}
We set $\mu_i = \mu_{b^*}$ and $\zeta_i = \mu_{b^*}^2$, as this choice of $\zeta_i$ is the natural one by the elimination of $\ln \, \mu_i^2/\zeta_i$ from the Wilson coefficients.
The $\otimes$ refers to the Mellin convolution on $x$.
Moreover, the perturbative Sudakov factor for the initial-state gluon is the following
\begin{equation}
    S_A(b_T;\mu_H,\mu_H^2) = \int^{\mu_H}_{\mu_{b^*}} \frac{d \bar{\mu}}{\bar{\mu}}\left[\Gamma_{cusp}(\bar{\mu})\, \ln\left(\frac{\mu_H^2}{\bar{\mu}^2}\right) + \gamma_V(\bar{\mu}) \right] + \mathcal{D}_{pert}(b_T^*;\mu_{b^*}) \ln \left(\frac{\mu_H^2}{\mu_{b^*}^2}\right)  \; ,
\end{equation}
and the NP parametrization is defined as
\begin{equation}
\begin{gathered}
    S_{\text{NP}}(x,b_T) = \mathcal S_{\text{NP}}(x,b_T) + \mathcal D_{\text{NP}}(b_T) \; ,\\
    \mathcal S_{\text{NP}}(x,b_T) = B_f(x)\, b_T^2 \quad \text{and} \quad \mathcal D_{\text{NP}}(b_T) = A_f \, \ln\left(\frac{\mu_H^2}{\mu_{\text{NP}}^2} \right) \, b_T^2 \; .
\end{gathered}
\end{equation}
The specific values of $B_f(x)$ depending on $x$ are shown in Ref.~\cite{Bor:2022fga}.
Also, $\mu_{\text{NP}}$ sets the separation between the perturbative and the NP behavior of the TMDPDF.

Having defined the TMDPDF and the TMDShF, we go through the convolution between both.
For the unpolarized $J/\psi$ leptoproduction, it simplifies to a product in $b_T$-space:
\begin{equation}\label{eq:conv}
\begin{aligned}
    \mathcal{C}\left[f_1^g S_{[n] } \right] & = \int_0^\infty \frac{d b_T}{2\pi} \, b_T \, J_0(b_T q_T)
    \times \sum_{[m]} C_{[m]}^{[n]}(b_T^*;\mu_{b^*},\zeta_{B,b}) \frac{\braket{\mathcal{O}^{[m]}}(\mu_{b^*})}{N_{pol}^{(J)}} \\
    \times & \sum_{a=q,\bar{q},g} \int_{x_A}^1 \frac{d\hat{x}}{\hat{x}} C_{g/a}(x_A/\hat{x}, b_T^*; \mu_{b^*}, \mu_{b^*}^2)\, f_{a/N}(\hat{x}, \mu_{b^*}) \\
    \times & \exp\left\{ -\int_{\mu_{b^*}}^{\mu_H} \frac{d \bar{\mu}}{\bar{\mu}} \left[\Gamma_{cusp}(\bar{\mu})\, \ln\left(\frac{\mu_H^2}{\bar{\mu}^2}\right) + \gamma_V(\bar{\mu}) + \gamma_s(\bar{\mu}) \right] \right\} \left( \frac{\mu_H^2}{\mu_{b^*}^2 \zeta_{B,b}}\right)^{- \mathcal{D}_{pert}(b_T^*;\mu_{b^*})} \\
    \times & \exp\left\{ -  \left[A_f\, \ln\left( \frac{\mu_H^2}{\mu_{\text{NP}}^2} \right) + A_S \, \ln\left( \frac{1}{\zeta_{B,0}}\right)+ B_f(x_A) + B_S \right] b_T^2 \right\} \; ,
\end{aligned}
\end{equation}
where $J_0$ is the zeroth first kind Bessel function. 
In this equation, we see the explicit effect of the TMDShF, where the dependence of the convolution, or the cross-section, on its perturbative and NP parameters is clear.

\begin{figure}[t]
    \centering
    \includegraphics[width=\textwidth]{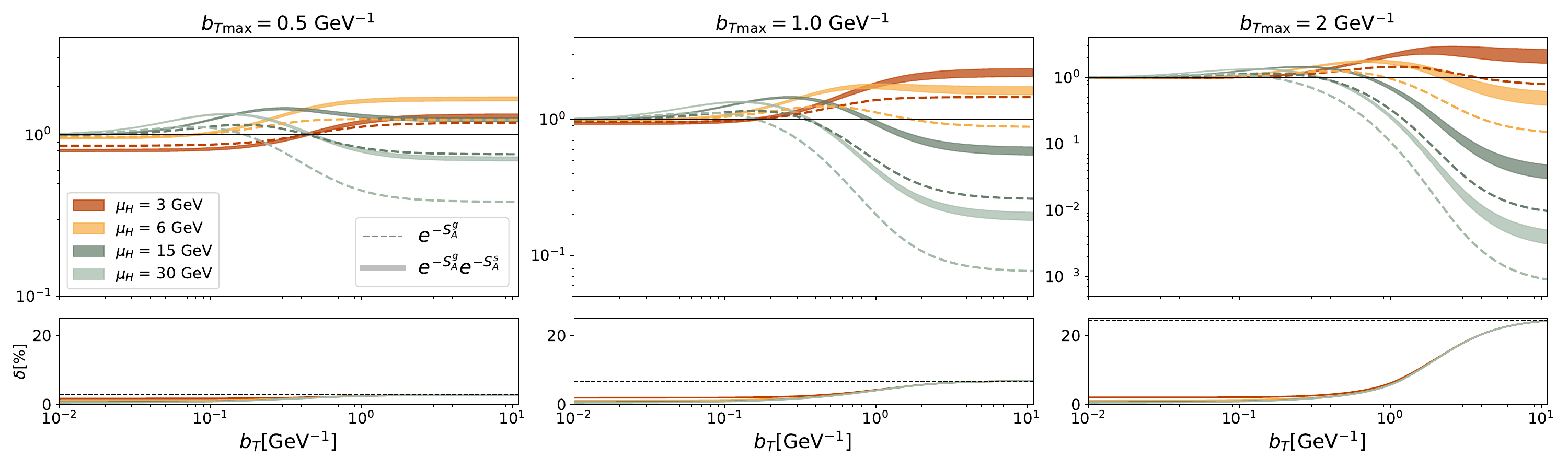}
    \caption{$b_T$-spectrum of $e^{-S_A^g}$ and of the third lane of Eq.~(\ref{eq:conv}) for several values of $\mu_H$ and for $b_{T\text{max}} = [0.5,1,2]$ GeV$^{-1}$. Bands are obtained by varying $\zeta_{B,b}$ by a factor of 2.}
    \label{fig:SA_pert}
\end{figure}

In Fig.~\ref{fig:SA_pert}, we illustrate the impact of the TMDShF on the TMD evolution of the convolution.\footnote{
It should be noted that a superscript $g$ has been introduced to indicate the evolution of the TMD gluon distribution, while a superscript $S$ denotes the evolution of the TMDShF.
}
Specifically, we show $e^{-S_A^g}$ at NNLL accuracy together with the third lane of Eq.~(\ref{eq:conv}), evaluated with $\gamma_s$ at NLO.
The uncertainty bands reflect the variation of the parameter $\zeta_{B,b}$ by a factor of two around its default value, $\zeta_{B,b}=1$.
Since the evolution depends on the $b_T^*$-prescription, and there is no unique choice for $b_{T\text{max}}$, we present results for three representative values.
The choice $b_{T\text{max}}=0.5$~GeV$^{-1}$ is clearly too small for $\mu_H=3,6$~GeV, as the distribution falls below unity already for $b_T \leq b_{T,\text{min}}$.
Conversely, $b_{T\text{max}}=2$~GeV$^{-1}$ lies beyond the region where the strong coupling remains perturbative; nevertheless, it is instructive to explore at which point the exponential factor for $\mu_H=3$~GeV starts decreasing within the perturbative domain.
Overall, we find that the effect of the TMDShF on the Sudakov factor of the convolution is to shift the distribution upward and toward larger $b_T$ values.
This occurs because $\gamma_s < 0$, thereby flipping the sign of the exponential.
Furthermore, the relative scale uncertainty, generally defined as $\delta = (f_{max} - f_{min})/2f_c$ with $f_{max/min}$ the maximum and minimum values of the envelope, and $f_c$ the central value, associated to a 3-point variation of $\zeta_{B,b}$ exhibits a universal behavior across energies and $b_{T\text{max}}$.
It grows rapidly at small $b_T$ and saturates at maximum values of 2.8, 6.8 and 24.3\% for $b_{T\text{max}}=$ 0.5, 1 and 2~GeV$^{-1}$, respectively.

\begin{figure}[t]
    \centering
    \includegraphics[width=\textwidth]{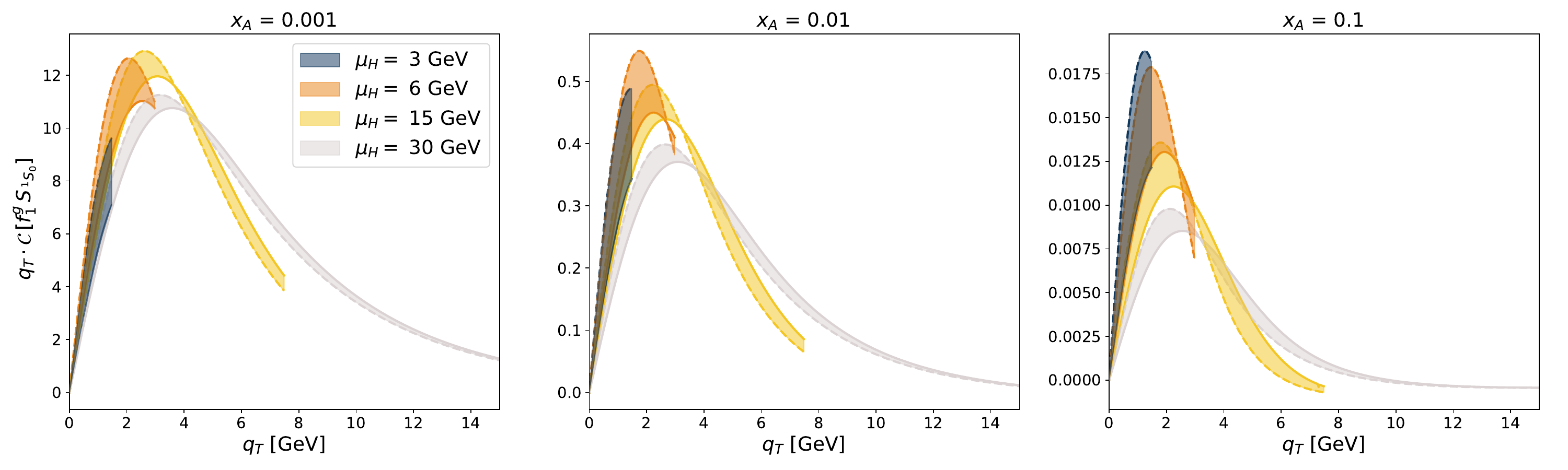}
    \caption{$q_T$-spectrum of $q_T \cdot \mathcal{C}[f_1^g S_{\,^1S_0^{[8]}}]$ for several values of the hard scale and for $x_A = [10^{-3}, 10^{-2}, 10^{-1}]$. Bands are obtain by varying the NP parameters of the TMD shape function.
    We plot for $b_{T\text{max}} = 1$ GeV$^{-1}$.}
    \label{fig:conv_qt}
\end{figure}

To explore the impact of the TMDShF in the NP region, we show in Fig.~\ref{fig:conv_qt} the quantity $q_T \cdot \mathcal{C}[f_1^g S_{^1S_0^{[8]}}]$, interpreted as a probability distribution, at the highest level of accuracy considered.
Results are presented for $A_S = B_S = 0$ (dashed curves) and for $A_S = 0.16$~GeV$^2$, $B_S = 0.5$~GeV$^2$ (solid curves).
The remaining parameters are fixed to $A_f = 0.16$~GeV$^2$, $\mu_{\text{NP}} = 1$~GeV, $\zeta_{B,0} = 0.1$, and $\zeta_{B,b} = 0.5$.
For the parton distribution functions and LDMEs we adopt the MSTW2008lo and SV sets, respectively.
We restrict the plots to $q_T \leq \mu_H/2$, which we identify as the region of validity of TMD factorization.
The convolution exhibits a strong dependence on the gluon PDF, and we therefore display results for several values of $x_A$.
The characteristic behavior of the gluon PDF is observed: for $x_A=0.1$, the peak of the distribution grows as the energy decreases.
At lower values of $x_A$, however, the peaks for $\mu_H=3$ and $6$~GeV approach those of higher energies, reflecting the faster fall-off of the gluon PDF at small $x_A$ and low scales—most prominently for $x_A=10^{-3}$.
Another general consequence of gluon PDF behavior is that, for any fixed $\mu_H$, the peak height decreases with increasing $x_A$.
Finally, the uncertainty bands represent the effect of the TMDShF in the non-perturbative region, evaluated for $(A_S,B_S)=(0.16,0.1)$~GeV$^2$, corresponding to values discussed in the previous section.
As expected, the width of the bands decreases with increasing $\mu_H$, since the exponential suppression of the NP model becomes stronger at larger hard scales.

\subsection*{Considerations for numerical results}

The complete and consistent expression for the convolution between a TMDPDF and a TMDShF in $J/\psi$ leptoproduction is given by Eq.~(\ref{eq:conv}).
Nevertheless, in line with the considerations discussed in this section, we adopt a set of simplifying assumptions for the numerical predictions of the cross-sections presented in the following.
In particular, the last term of Eq.~(\ref{eq:conv}) defines the non-perturbative model of the convolution, which remains an undetermined component in our analysis.
Here the $B$-terms govern the large-$b_T$ behavior of the TMDs and must be treated as independent parameters.
By contrast, the $A$-terms share the same physical origin, namely the contribution of soft-gluon radiation to the process in the NP regime.
Since $A_S$ and $A_f$ naturally lie within the same range of values (see discussion in connection with Fig.~\ref{fig:TMDShF-qt}), we impose the identification
\begin{equation}
    A \equiv A_S = A_f \; .
\end{equation}
Consequently, the factor $\ln(\mu_H^2 / \mu_{\text{NP}}^2 \zeta_{B,0})$ appears multiplied by $A$.
Further, since both $\mu_{\text{NP}}$ and $\zeta_{B,0}$ are arbitrary parameters delineating the perturbative–non-perturbative separation of the two TMDs, it is convenient to introduce a combined NP parameter,
\begin{equation}
    \tilde{\mu}_{\text{NP}}^2 \equiv \mu_{\text{NP}}^2 \, \zeta_{B,0} \; .
\end{equation}
In this way, the non-perturbative model for the convolution of a TMDPDF and a TMDShF can be expressed as:
\begin{equation} \label{eq:SNP-lp}
    S^{\ell p}_{\text{NP}} = \left[A \,\ln\left(\frac{\mu_H}{\tilde{\mu}_{\text{NP}}}\right) + B_f(x_A) + B_S \right]  b_T^2  \; ,
\end{equation}
where the superscript $\ell p$ highlight that this is the NP model for the leptoproduction cross-section.
While the above parametrization offers a simple and practical framework for exploring the low–transverse-momentum region, where the TMDShF plays a significant role, a more refined modeling of $S_{\text{NP}}$—or of the $b_T$-prescription employed to separate the perturbative from the non-perturbative domain—will ultimately be required for precision phenomenology.
For the present study, and in line with global fits to low-energy SIDIS data as well as high-energy Drell–Yan and $Z$-boson production data, we adopt $A=0.414$~GeV$^2$ and $\tilde{\mu}_{\text{NP}}=1.6$~GeV as central values, and vary $A$ within the range discussed above when presenting the predictions of the next section. Whereas, $B_f(x)$ is parameterized as
\begin{eqnarray}
    B_f(x)=\frac{9}{8}(-g_1\log(2)+g_2(1+2g_3\log(\frac{10xx_0}{x_0-x})))
\end{eqnarray}
with $g_1=0.184~\text{GeV}^2$, $g_2=0.201~\text{GeV}^2$, $g_3=-129~\text{GeV}^2$ and $x_0=0.009$ \cite{Aybat:2011zv}. A point to note is that the parameters were obtained for the case of quark TMDs (two); hence, to use them for the gluon TMD, an additional factor of $9/8(\equiv C_A/(2C_F))$ is multiplied. 

\section{Theoretical predictions} \label{sec:theoretical-predictions}

\begin{table}[t]
    \centering
    \scriptsize
    \begin{tabular}{|l|l|l|l|}
    \hline
    \textbf{Element} & \textbf{Symbol} & \textbf{Comment} & \textbf{Reference} \\
    \hline
    \rowcolor{gray!15}
    \multicolumn{4}{|c|}{\textbf{General}} \\
    \hline
    Charm quark mass & $m_c$ & $m_c = 1.50$ GeV &  \\
    $J/\psi$ mass & $M$ & $M = 2m_c=3$ GeV & \\
    Proton mass & $M_N$ & $M_N \approx 0$ GeV & \\
    Photon's virtuality & $Q^2$ & $q^2 = -Q^2$, $q$ photon's momentum & \\
    QED coupling const. & $\alpha_{em}$ & $\alpha_{em} = 1/137$ &  \\
    QCD coupling const. & $\alpha_s \equiv 4\pi \, a_s$ & $\alpha_s(M) \simeq 0.261$ taken at three-loop & \cite{ParticleDataGroup:2024cfk} \\
    WW function & $f_{\gamma/e}$ & $Q_{min}^2 = m_e^2 y^2/(1-y), Q_{max}^2 = 2.5 GeV^2$
    & \cite{Frixione:1993yw} \\
    Dirac gamma 5 & $\gamma_5$ & Larin scheme in $d$-dimensions & \\
    \hline
    \rowcolor{gray!15}
    \multicolumn{4}{|c|}{\textbf{Hadron tensor}} \\
    \hline
    [n] Hard function & $H_{[n]}$ & $\mathcal{O}(a_s^1)$ & This paper \\
    Cusp AD & $\Gamma_{cusp}$ & $\mathcal{O}(a_s^3)$ & \cite{Vladimirov:2017ksc} \\
    TMDPDF AD & $\gamma_V$ & $\mathcal{O}(a_s^2)$ & \cite{Echevarria:2015uaa} \\
    TMDShF AD & $\gamma_s$ & $\mathcal{O}(a_s^1)$ & \cite{Echevarria:2024idp} \\
    Factoriz. scales & $\mu_H, \zeta_A, \zeta_B$ & $\mu_H = (M^2+Q^2)/M$, $\zeta_A = \mu_H^2$, $\zeta_B = 1$ &  \\
    Natural scales & $\mu_b,\, \zeta_b$ & $\mu_b = b_0/b_T$, $\zeta_{b} = \mu_b^2$ &  \\
    \hline
    \rowcolor{gray!15}
    \multicolumn{4}{|c|}{\textbf{Collins-Soper kernel} $\mathcal D$} \\
    \hline
    Pert.~part & $\mathcal{D}_{pert}$ & $\mathcal{O}(a_s^2)$ &  \cite{Echevarria:2012pw} \\
    NP part & $\mathcal{D}_{\text{NP}}$ & 1 parameter: $A$ &  \\
    Defining scale & $\mu_{b^*}$ & $\mu_b^*\equiv b_0/b_T^*$ where, $b_0=2e^{-\gamma_E}$ &  \\
    $b_T$-prescription & $b_T^*$ & $b_{T\text{max}} = 1.5$ GeV$^{-1}$ & \\
    \hline
    \rowcolor{gray!15}
    \multicolumn{4}{|c|}{\textbf{Unpolarized gluon TMDPDF $f_{1}^g$}} \\
    \hline
    Matching coef. & $C_{f \leftarrow f'}$ & $\mathcal{O}(a_s^1$) & \cite{Echevarria:2015uaa} \\
    Unpol. PDF & $f_1$ & NNLO from MSTW2008 PDF set & \cite{Martin_2009}\\
    Log.~fraction momentum & $x_A=x$ & Eq.~(\ref{eq:x}) &\\
    NP part & $\mathcal{S}^g_{\text{NP}}$ & 1 parameter: $B_f(x_A)$ &  \\
    OPE scale & $\mu_{OPE} = \mu_b$ & Central value & \\
    \hline
    \rowcolor{gray!15}
    \multicolumn{4}{|c|}{\textbf{Unpolarized $[n]$-TMDShF $S_{[n]}$}} \\
    \hline
    Matching coef. & $C_{[n]}^{[m]}$ & $\mathcal{O}(a_s^1)$ & \cite{Echevarria:2024idp} \\
    \text{[n]}LDME & $\braket{\mathcal{O}_{[n]}}$ & We use several sets (see in the text) & \cite{brambilla2024, Chao_2012, Sharma_2013, Bodwin_2014}  \\
    NP part & $\mathcal{S}^s_{\text{NP}}$ & 1 parameter: $B_S$ &  \\
    OPE scale & $\mu_{OPE} = \mu_b$ & Central value &  \\ 
    \hline
    \end{tabular}
    \caption{Summary of the theory necessary for the calculation.}
    \label{tab:summary_symbols}
\end{table}

In this section, we present theoretical predictions for the transverse-momentum-differential cross-section of unpolarized $ J/\psi $ electroproduction at low and intermediate transverse momentum, based on the TMD factorization formalism established in Section~\ref{sec:TMDShF}. These predictions are particularly relevant for future measurements at the Electron-Ion Collider (EIC), where the NP transverse dynamics of the gluon TMD and soft gluon radiation can be studied with high precision.

Our predictions are obtained within the TMD factorization framework, working at leading power in the $P_{\psi\perp}/Q$ expansion.
Large logarithms are resummed to NNLL$^\prime$ accuracy, and the perturbative hard function is included at $\mathcal{O}(\alpha_s)$ with full one-loop virtual corrections.
The analysis is restricted to the color-octet channel, which is expected to dominate in the inelastic regime at large photon virtuality, $Q^2 \gg M^2$.
By contrast, the color-singlet contribution appears only at $\mathcal{O}(\alpha_s^2)$ and is suppressed in the strong coupling, while diffractive contributions at $z=1$ are further power suppressed by $1/Q^2$ relative to the CO channel~\cite{Fleming:1997fq}.

Given the number of parameters involved, and in order to maintain clarity, we summarize in Tab.~\ref{tab:summary_symbols} the notation and input required for the theoretical calculation.
The theoretical formulas are shown in Sec.~\ref{sec:csTMD} and Sec.~\ref{sec:TMDShF}, but we emphasize that the predictions have been obtained in the photon frame, where the observed transverse momentum $\bm{P}_{\psi\perp}$ is directly sensitive to the intrinsic transverse momentum of the incoming gluon.
However, the factorization formula in Eq.~(\ref{eq:Hadronic-Tensor}) is defined in the hadron frame, where the transverse momentum variable is the transverse momentum of the virtual photon, $\bm{q}_{\perp}$.
Both are related via
\begin{equation}
    \bm{q}_\perp=\frac{\bm{P}_{\psi\perp}}{z} \; ,
\end{equation}
where $z\sim 1$, as enforced by the delta function in Eq.~(\ref{eq:Hadronic-Tensor}),
and in this frame, the leading-twist collinear momentum fraction of the parton is given by 
\begin{eqnarray}\label{eq:x}
    x=x_B\frac{M^2+Q^2}{Q^2} = \frac{M^2+Q^2}{y \,S} \; .
\end{eqnarray}


\begin{figure}[t]
    \centering
		\begin{subfigure}{0.49\textwidth}
        \centering
		\includegraphics[height=5cm,width=7cm]{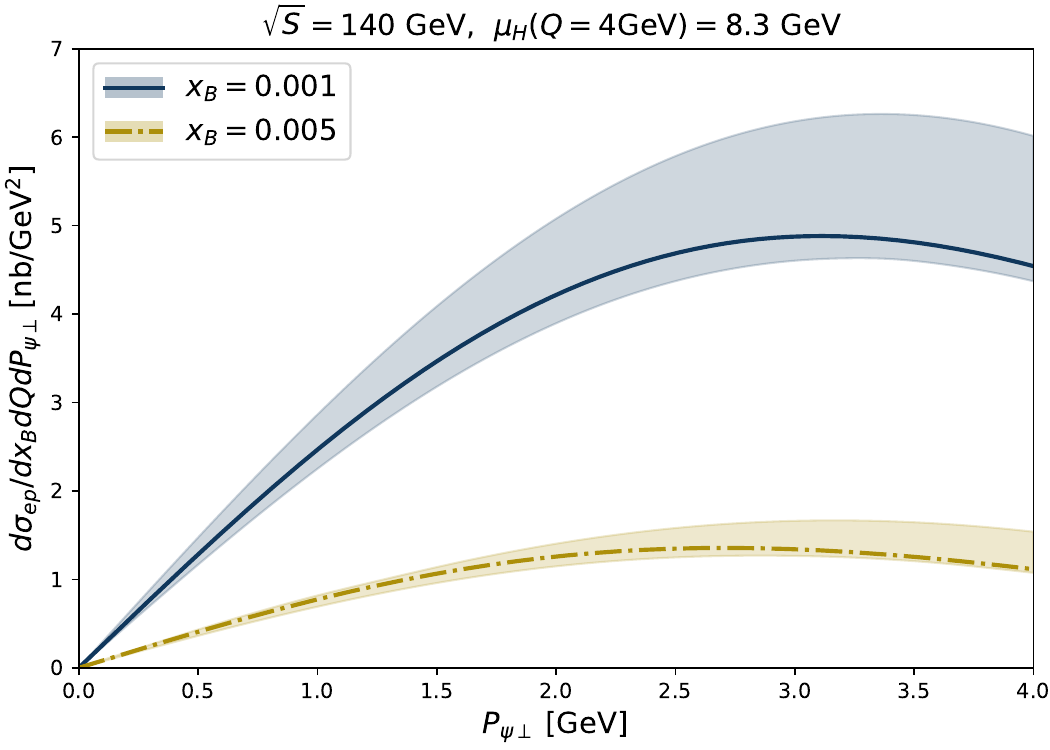}
			\caption*{(a)}
		\end{subfigure}
	    \begin{subfigure}{0.49\textwidth}
        \centering
	    \includegraphics[height=5cm,width=7cm]{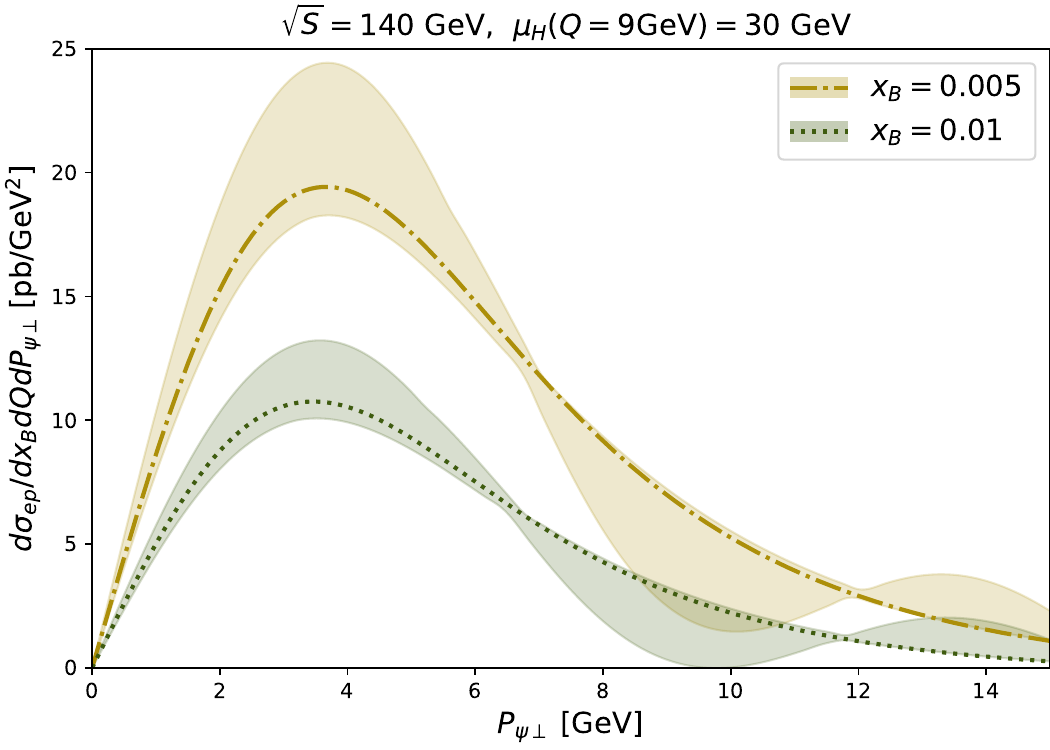}
	        \caption*{(b)}
	    \end{subfigure}
        \vfill
        \centering
		\begin{subfigure}{0.49\textwidth}
        \centering
		\includegraphics[height=5cm,width=7cm]{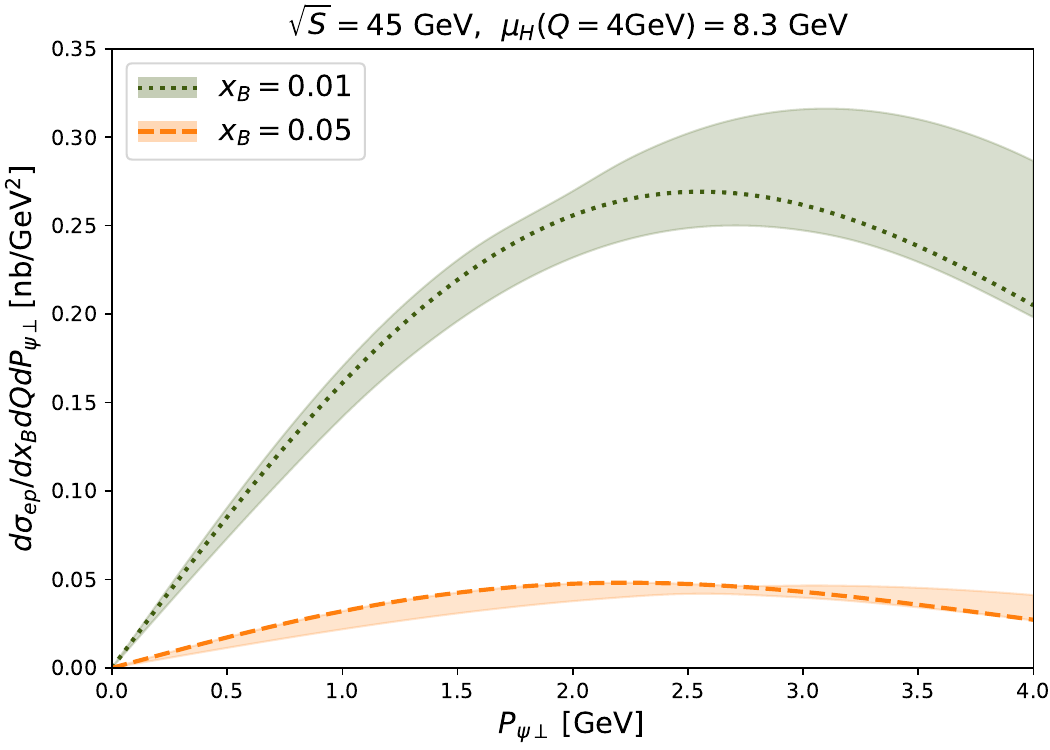}
			\caption*{(c)}
		\end{subfigure}
	    \begin{subfigure}{0.49\textwidth}
        \centering
	    \includegraphics[height=5cm,width=7cm]{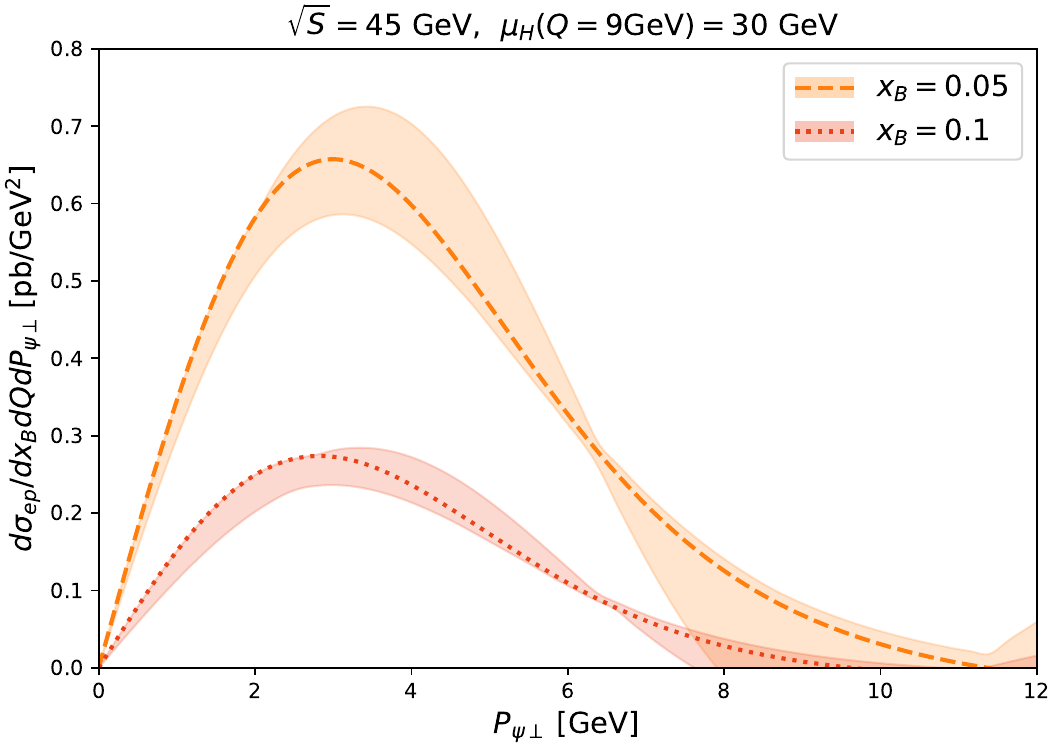}
	        \caption*{(d)}
	    \end{subfigure}    
        \vspace{-0.5em}
	\caption{Differential cross-section as functions of $J/\psi$ transverse momentum, $d\sigma_{e p}/dP_{\psi\perp}$ at $\sqrt{s}=140$ GeV (a,b) and $\sqrt{s}=45$ GeV (c,d) for various values of $Q$ and $x_B$. The band represents an independent variation in both the renormalization scales $\mu_0$ and the rapidity scale $\zeta_0$ by a factor of 2 around their default input scale.
    \label{fig:cs_eic_electro_scale}}
\end{figure}

For the theoretical estimates, we choose two extreme center-of-mass energy, $\sqrt{S}=140$ and $45$ GeV. We show the $P_{\psi\perp}$ spectrum for $P_{\psi\perp}<\mu_H/2$, a reliable TMD factorization domain. We used MSTW2008 at NNLO PDF set~\cite{Martin:2009iq}.
The minimum scale of the PDF is 1 GeV.

In Fig.~\ref{fig:cs_eic_electro_scale}, we present the $P_{\psi \perp}$-spectrum of the differential cross-section for $ \sqrt{s} = 140\,\text{GeV} $ (a,b) and  $ \sqrt{s} = 45\,\text{GeV} $ (c,d) for different values of $ x_B $ and $Q = 4, 9$ GeV.
We observe a peaked structure in the spectrum, with the position of the peak shifting toward larger $P_{\psi\perp}$ and becoming narrower as $Q$ increases which is consistent with expectations from resummed TMD calculations as the behavior arises from stronger Sudakov suppression at higher scales, and from the width of the NP effect.
Moreover, we observe a significant change in the profile of the distribution in the TMD region when comparing the plots for $Q = 4$ GeV and $Q = 9$ GeV.
This behavior is attributed to the fact that the hard scale $\mu_H$ increases quadratically with $Q$, and consequently, the condition $P_{\psi \perp} < \mu_H/2$, under which our prediction is reliable, covers almost the entire $P_{\psi \perp}$-spectrum where the cross-section is not zero.
This result may indicate that the energy $Q = 9$ GeV is too large for our factorization framework, potentially requiring (at this energy and above), a proper scale separation between $Q$ and $M$. In other words, a specific TMD factorization approach valid for $Q \gg M$ may be necessary.
We leave this analysis to future work.
To conclude with this figure, we include uncertainty bands arising from variations of the rapidity and renormalization scales by a factor of two. Although not explicitly shown here, we observe narrower uncertainty bands at higher resummation accuracy, as expected.
Moreover, one obtain the oscillatory scale uncertainty band at large $P_{\psi\perp}$ is the consequence of the minimum scale of the PDF being 1 GeV and higher.

\begin{figure}[t]
    \centering
		\begin{subfigure}{0.49\textwidth}
        \centering
		\includegraphics[height=5cm,width=7cm]{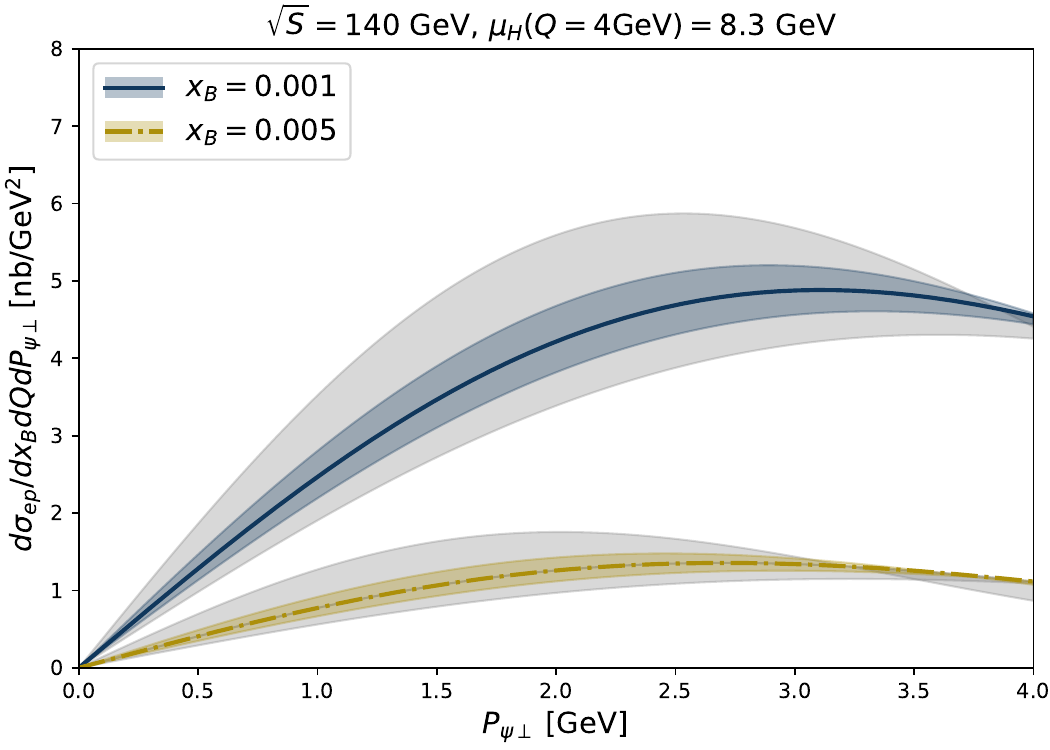}
			\caption*{(a)}
		\end{subfigure}
	    \begin{subfigure}{0.49\textwidth}
        \centering
	    \includegraphics[height=5cm,width=7cm]{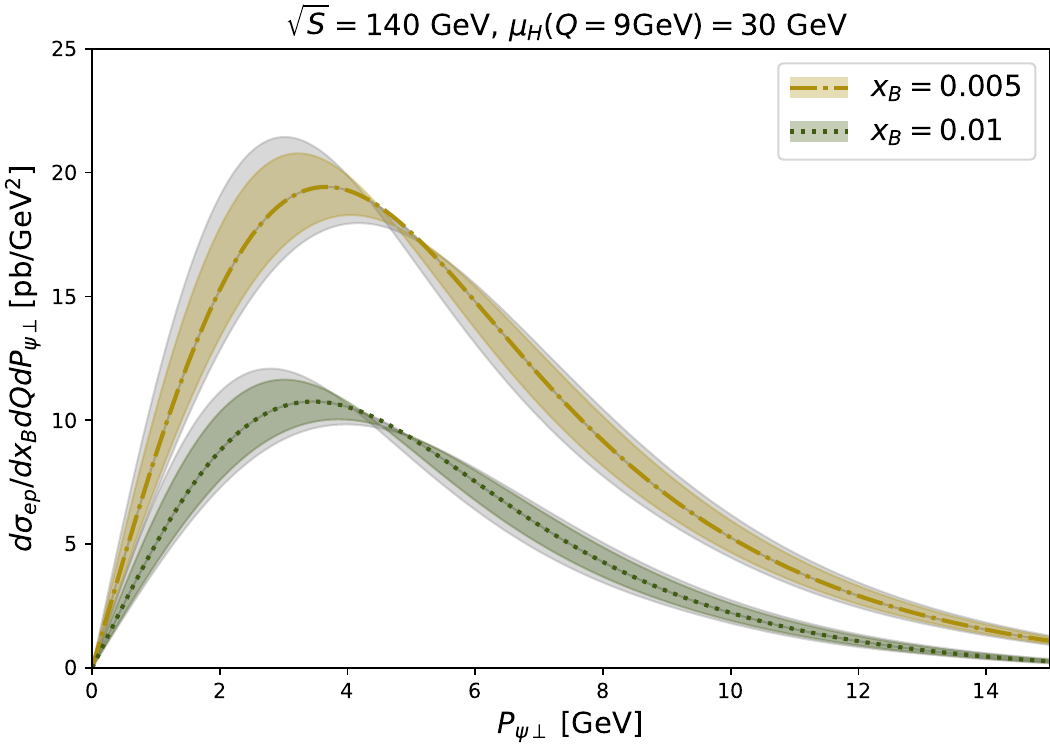}
	        \caption*{(b)}
	    \end{subfigure}
        \vfill
        \begin{subfigure}{0.49\textwidth}
        \centering
		\includegraphics[height=5cm,width=7cm]{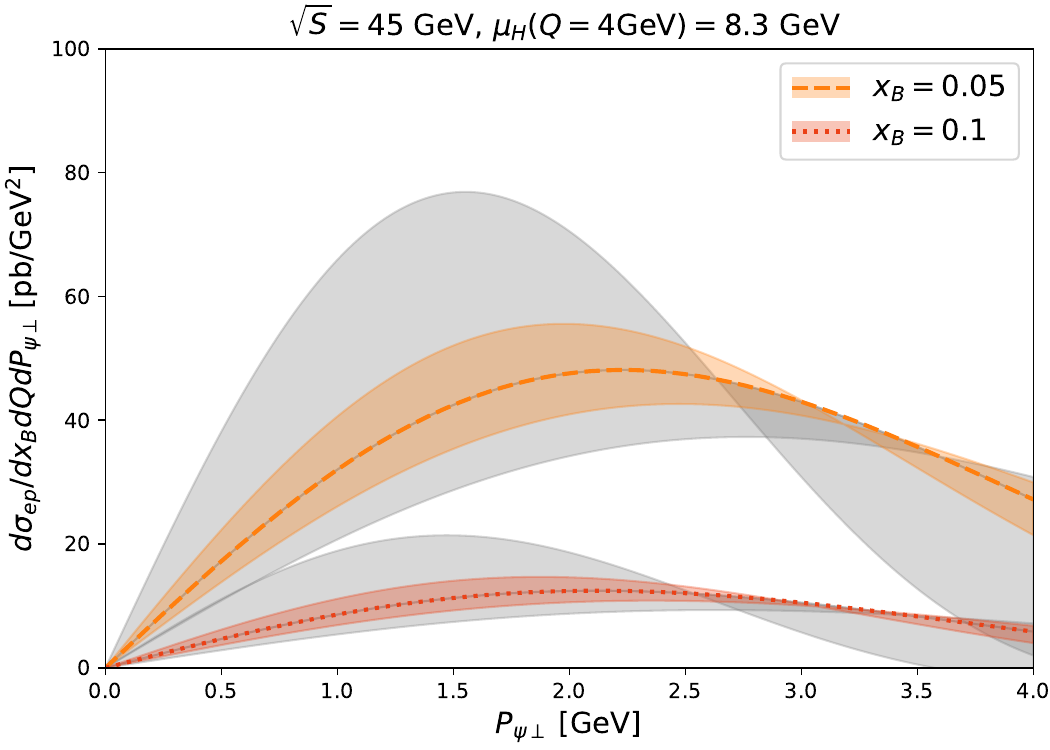}
			\caption*{(c)}
		\end{subfigure}
	    \begin{subfigure}{0.49\textwidth}
        \centering
	    \includegraphics[height=5cm,width=7cm]{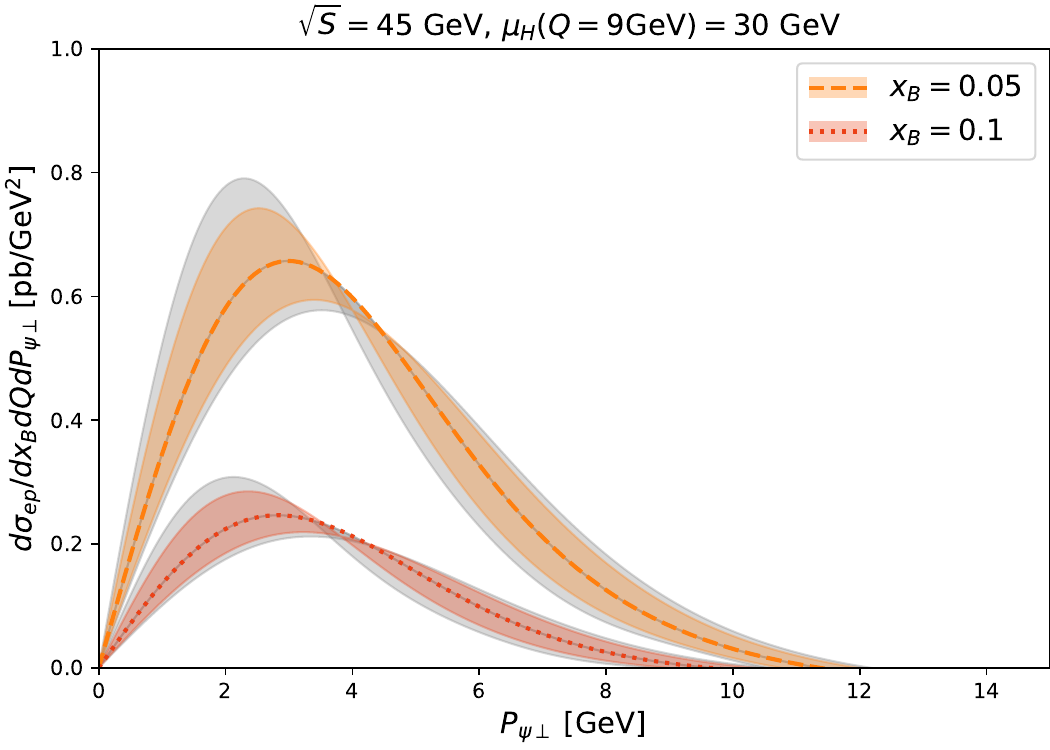}
	        \caption*{(d)}
	    \end{subfigure}
        \vspace{-0.5em}
	\caption{Differential cross-section as functions of $J/\psi$ transverse momentum $d\sigma_{e p}/dP_{\psi\perp}$ with uncertainty band due to NP effect. The inner band corresponds to variation in $ A \in [0.05, 0.8]\,\text{GeV}^2 $, and the outer band to $ B_S \in [0, 3]\,\text{GeV}^2 $.    
    \label{fig:cs_eic_electro_snp}}
\end{figure}

\begin{figure}[t]
    \centering
		\begin{subfigure}{0.49\textwidth}
        \centering
		\includegraphics[height=5cm,width=7cm]{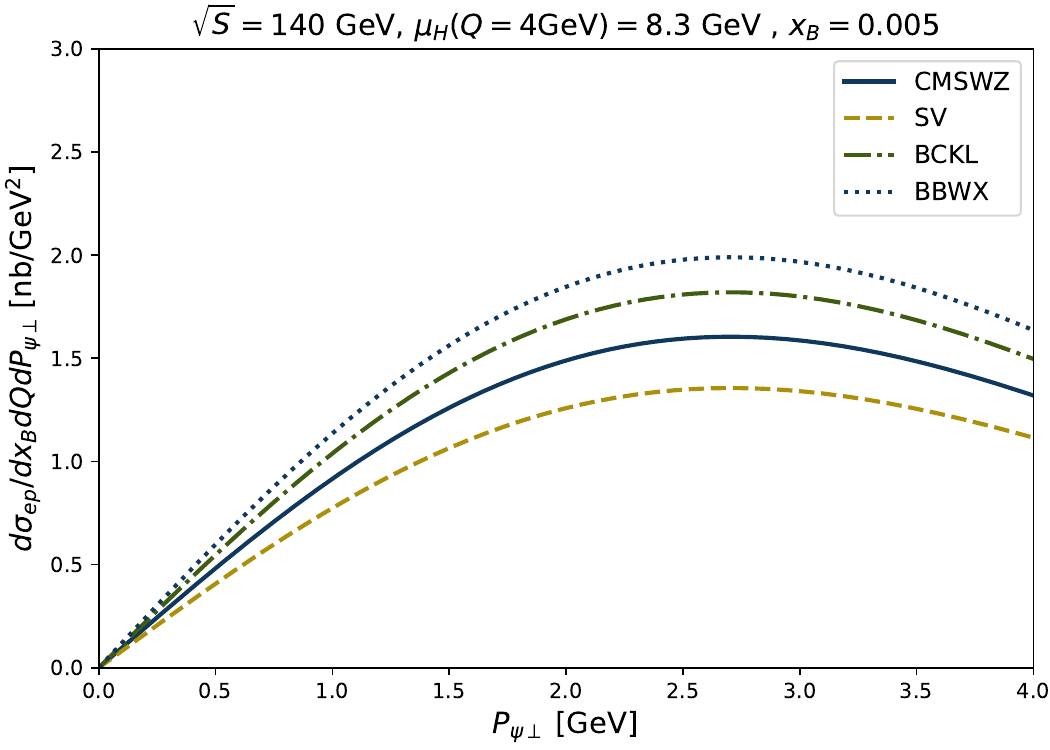}
			\caption*{(a)}
		\end{subfigure}
	    \begin{subfigure}{0.49\textwidth}
        \centering
	    \includegraphics[height=5cm,width=7cm]{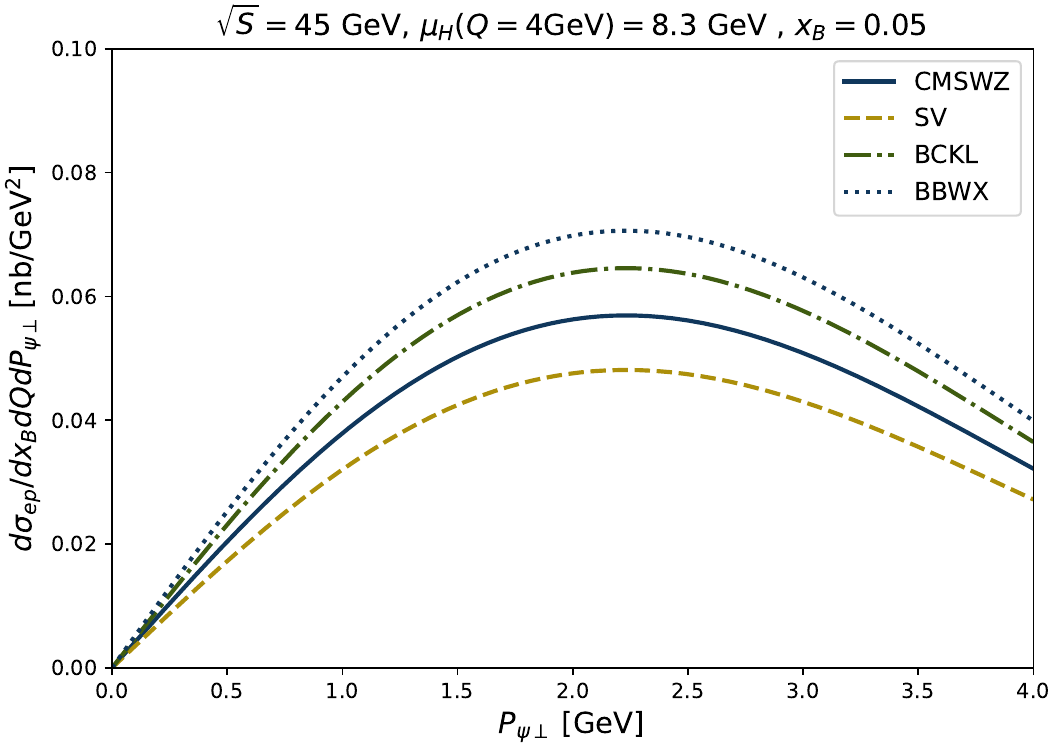}
	        \caption*{(b)}
	    \end{subfigure}
        \vspace{-0.5em}
	\caption{Differential cross-section as functions of $J/\psi$ transverse momentum $d\sigma_{e p}/dP_{\psi\perp}$ for different sets of LDMEs: BBXW\cite{brambilla2024}, CMSWZ\cite{Chao_2012}, SV\cite{Sharma_2013}, and BCKL\cite{Bodwin_2014}.
    \label{fig:cs_eic_electro_ldme}}
\end{figure}

\begin{figure}[t]
    \centering
    \includegraphics[width=0.33\textwidth]{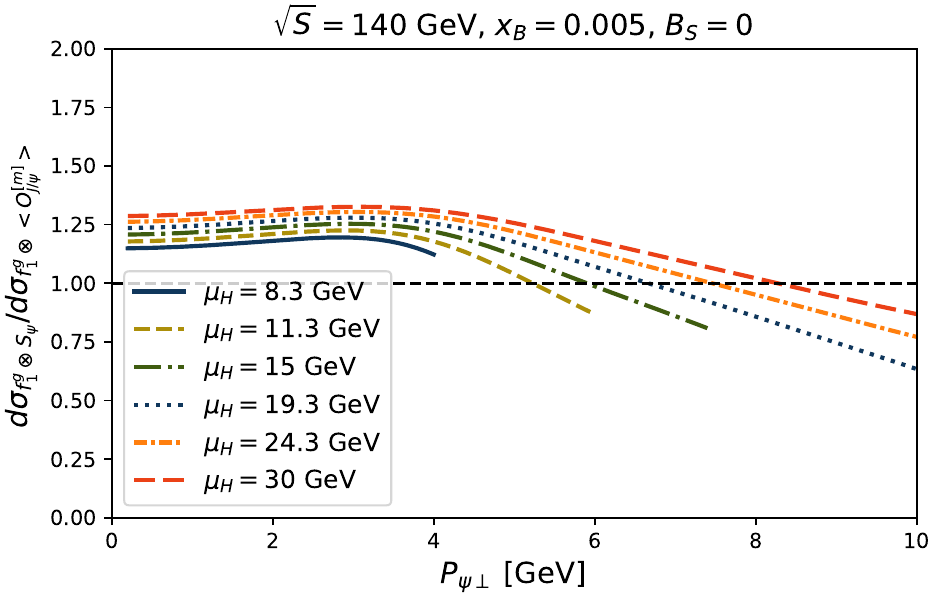}
    \includegraphics[width=0.31\textwidth]{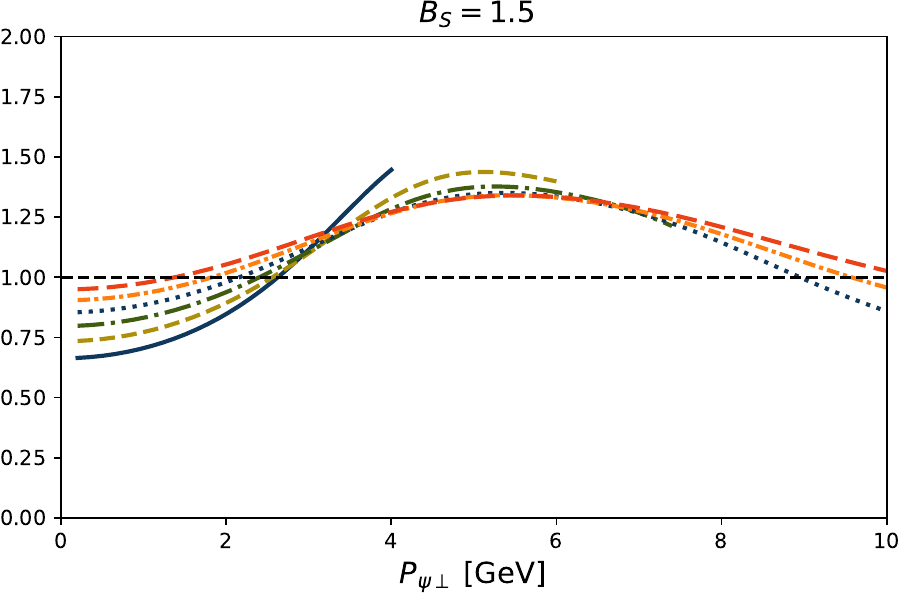}
    \includegraphics[width=0.31\textwidth]{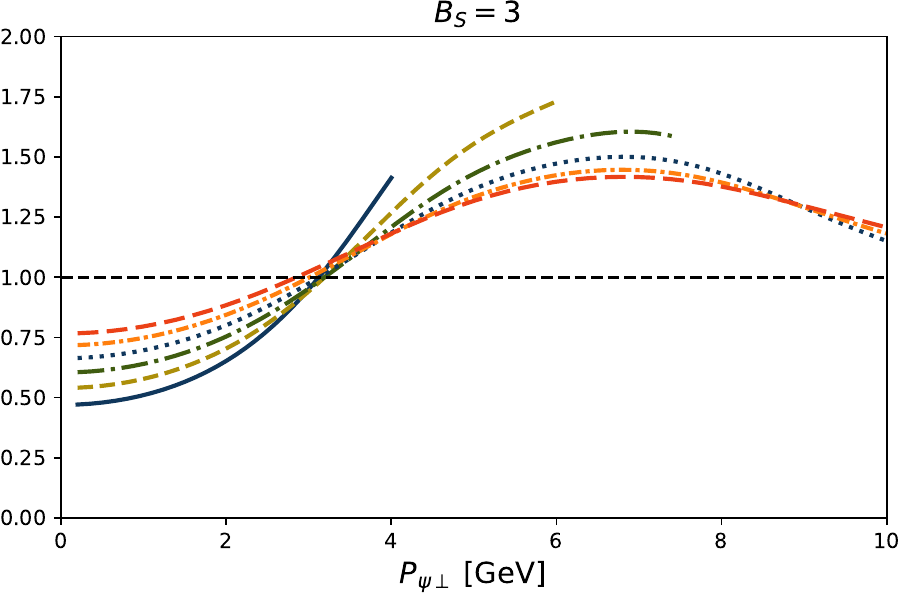}
    \caption{Ratio $ R(P_{\psi\perp}) $ given by Eq.~(\ref{eq:ratio_ShF-NRQCD}) at $ \sqrt{s} = 140\,\text{GeV} $, $ x_B = 0.005 $, for $B_S = 0$~GeV$^2$ (left), $B_S = 1.5$~GeV$^2$ (center), and $ B_S = 3$~GeV$^2$ (right).}
    \label{fig:ratio-plot}
\end{figure}

In Fig.~\ref{fig:cs_eic_electro_snp} we examine the sensitivity of our predictions to the NP modeling of the TMD convolution, Eq.~(\ref{eq:SNP-lp}), so the uncertainty band is generated by varying both the parameters $A$ and $B_S$. 
The lines show the so-called central prediction which is defined by the above discussed value $A = 0.414$ GeV$^2$ and $B_S = 1.5$ GeV$^2$.
The inner band spans the range $0.05 \leq A \, \text{GeV}^{-2} \leq 0.8$, which defines a variation in the effective transverse distance at the corresponding hard scale $\mu_H $, and for $\tilde{\mu}_{\text{NP}} = 1.6$ GeV.
According to the discussion below on $B_S$, we take the range $0 \leq B_S \, \text{GeV}^{-2} \leq 3$ which is shown by the gray bands.
As we saw in Fig.~\ref{fig:TMDShF-energy_BS}, the smaller the value of $B_S$, the greater the maximum value of the distribution at low transverse momentum, so the upper limit of the band corresponds to $B_S = 0$ GeV$^2$.
Based on the width of the uncertainty bands arising from the NP parameters, it is evident that parameter $B_S$ has a more significant impact on the cross-section.
Moreover, as expected, the uncertainty associated with $B_S$ decreases with increasing $Q$, as can be seen by comparing it with the uncertainty from $A$ in (b) and (d).

Moreover, we obtain a negative (unphysical) cross-section as $P_{\psi\perp}\to\mu_{H}/2$, particularly for large $Q(9~\text{GeV})$ at large $x$. This is due to the behavior of TMD convolution in that domain; see Fig.~\ref{fig:conv_qt}. Such behavior of a negative cross-section has also been reported in Ref~\cite{Boer:2023zit}. However, with a change (increase) in the value of NP parameters, which means suppressing the non-perturbative contributions further, we can obtain the positive cross-section even at the large $P_{\psi\perp}\to\mu_H/2$. However, strictly speaking, the TMD factorization formalism is applicable for $P_{\psi\perp}<<\mu_H$. The appearance of negative values near $P_{\psi\perp}\sim\mu_H/2$ signals the breakdown of the TMD expansion near the (assumed) boundary of its applicability, rather than an inconsistency of the formalism within its domain of validity. 
Additionally, we do not have constraints on the non-perturbative parameters yet, particularly for the gluon TMD, from the experiments. This leads to the conclusion that a better parameterization is required, particularly for the parameter that depends on $x$; $B_f(x)$. Moreover, particularly at large $P_{\psi\perp}$, power correction and higher twist effect might play a sizable role. We left these analysis for future work.

The final analysis in our predictions comes due to the different LDME sets.
In Fig.~\ref{fig:cs_eic_electro_scale} and~\ref{fig:cs_eic_electro_snp}, we show the prediction using SV LDME set~\cite{Sharma_2013}. However, we have different sets of LDMEs in the literature, where some of the sets are extracted from global fit, but with high $P_T$ data. 
In particular, each line in Fig.~\ref{fig:cs_eic_electro_ldme} shows the NNLL$'$ prediction obtained using the following LDME sets: BBXW~\cite{brambilla2024}, CMSWZ~\cite{Chao_2012}, SV and BCKL~\cite{Bodwin_2014}
We see a wider spectrum using different LDMEs than in the theoretical uncertainty band, which tells a better fit is needed to extract them, especially in low-$P_T$.

\subsection{On the TMDShF influence}

In light of the results presented regarding the effects of the TMD shape function throughout this work, its inclusion may therefore play a significant role in improving the description of the data for the $\mathcal{O}(\alpha_s)$ color-octet channel at low-transverse-momentum.
These observations motivate a more detailed investigation of this particular impact of the TMDShF, as it appears to lead a $q_T$ spectrum which may either undershoot or overshoot the corresponding result obtained without its inclusion—that is, the standard NRQCD prediction.
This sensitivity is primarily governed by the parameter $B_S$, which controls the magnitude of the TMDShF contribution in the non-perturbative region, as well as by the choice of the hard scale $\mu_H$.

To disentangle these effects, we consider the ratio:
\begin{equation} \label{eq:ratio_ShF-NRQCD}
    R(P_{\psi\perp}) \equiv \frac{d\sigma_{f_1^g\otimes S_{\psi}}}{d\sigma_{f_1^g\otimes <O_{J/\psi}^{[m]}>}} \; ,
\end{equation}
which directly quantifies the suppression induced by the TMD shape function in the low-$ P_{\psi\perp} $ region.
Here $d {\sigma}_{f_1^g \otimes S_\psi}$ and $d\sigma_{f_1^g\otimes <O_{J/\psi}^{[m]}>}$ denote the differential cross-section with and without TMDShF, respectively.
The results are displayed in Fig.~\ref{fig:ratio-plot} for EIC kinematics, considering $B_S = 0,$ 1.5 and 3 GeV$^2$ and varying $\mu_H$, with the spectrum plotted up to $P_{\psi\perp} = \mu_H/2$.
We used SV LDMEs set for these plots. Moreover, the LDME sets have a negligible or no effect on its behavior. The reason lies in the perturbative tails, as they are the same for $S-$ and $P-$ states except for the cross terms, which we have considered zero in our numerical estimates as discussed in section~\ref{sec:TMDShF}. Hence, the ratio mostly represents the effect of the shape function.

We observe that, unlike the case $B_S \neq 0$, the prediction including the TMDShF overshoots the standard NRQCD result.
For the left panel, where the NP contribution of the TMDShF is effectively switched off, the ratio remains close to unity, and the residual variation (within about 20\%) originates mainly from scale dependence.
As $B_S$ increases, corresponding to an enhanced strength of the TMDShF, a clear distortion of the ratio develops.
For $B_S = 1.5$~GeV$^2$ exhibits a pronounced peak around $P_{\psi \perp} \sim 5-6$ GeV.
This indicates a moderate enhancement of the cross-section in this intermediate momentum region, likely arising from constructive interference or modified phase-space weighting due to the inclusion of the TMDShF.
The dependence on the scale becomes stronger in this regime.
For $B_S = 3$~GeV$^2$, the effect becomes even more pronounced.
The peak in the intermediate region increases, while at larger momenta the ratio falls below unity.
The transition point between enhancement and suppression is roughly stable across $\mu_H$, indicating a dynamical origin tied to the intrinsic shape of the additional function rather than to the perturbative scale choice.
Overall, the results demonstrate that the inclusion of the TMDShF governed by $B_S$ introduces nontrivial shape modifications to the transverse-momentum distribution.
Increasing $B_S$ amplifies these distortions, confirming its role as a strength or broadness parameter of the TMDShF.
We emphasize, however, that this behavior pertains to the specific implementation described in Eq.~(\ref{eq:SNP-lp}), and could change under alternative NP parametrization of the CS kernel and the gluon TMDPDF.

It is worth to emphasize that the present analysis has been confined to the case of electroproduction.
Nonetheless, the theoretical framework developed in this work can be straightforwardly extended to provide predictions for photoproduction processes.
It is, however, well established that the H1 measurements of large-$p_T$ $J/\psi$ photoproduction~\cite{H1:2010udv} are accurately reproduced by the color-singlet contribution alone, computed at $\mathcal{O}(\alpha_s^2)$~\cite{Flore:2020jau}, where the CO contribution at $\mathcal{O}(\alpha_s)$ is known to significantly overshoot the experimental data.
Moreover, theoretical predictions based on the conventional NRQCD factorization framework are known to fail in describing the data, particularly in the low-$p_T$ and high-$z$ region, as discussed in Ref.~\cite{Brambilla:2024iqg}.
In a preliminary analysis, as a qualitative comparison, we confront our estimate for photoproduction at $z \sim 1$ with the H1 measurement in the $0.75 < z < 0.9$ bin~\cite{H1:2010udv}.
We find that the incorporation of the TMDShF within the TMD framework leads to theoretical predictions of the CO channel that exhibit markedly better agreement with the experimental data, yielding lower cross sections at small $p_T$ relative to the current collinear NRQCD results.
However, we leave a detailed analysis of photoproduction in the TMD framework for future work.

\section{Conclusions} \label{sec:conclusion}

In this work, we have investigated the $J/\psi$ production in electron–proton scattering within the TMD factorization framework, with a particular focus on the role of the TMD shape function.
We have performed evolution at what we denoted as NNLL$'$ with the hard function computed at NLO.
We have performed a thorough investigation of the TMD evolution of the TMD shape function, including its impact on the perturbative and the non-perturbative parts of the convolution with the unpolarized TMD gluon distribution.

We have calculated the virtual one-loop contribution to the electroproduction cross section, thereby obtaining the hard function required for a more accurate analysis of the cross section in the low-$p_T$ region.
In this context, we have resolved the discrepancy present in previous studies concerning the TMD shape function at next-to-leading order, as well as its unphysical $Q^2$ dependence.

We further find that the impact of the TMDShF on the perturbative Sudakov factor of the convolution is to shift the distribution in impact-parameter space upward and toward larger $b_T$ values.
Concerning the low-$p_T$ cross-section predictions, for $Q = 4$~GeV, the non-perturbative effect of the TMDShF itself—parameterized by $B_S$—dominates over the contribution of the Collins-Soper kernel, as evidenced by the width of the uncertainty bands obtained when varying the corresponding NP parameters over a considerable range.
In contrast, for $Q = 9$~GeV, the effects of both contributions are found to be approximately comparable, as expected increasing $Q$.
We have also quantified the variation of our predictions arising from the choice of the LDME set.
This uncertainty is larger than that associated with the NP parametrization, highlighting the need for its reduction.

Furthermore, the analysis of the cross-section predictions indicates that a dedicated TMD factorization framework, valid in the regime $Q \gg M$, may be necessary.
A detailed investigation of this possibility is deferred to future work.

Finally, we have observed that the TMDShF plays a central role in governing the low-$p_T$ behavior of the cross-section.
We have performed a quantitative comparison for EIC kinematics with the standard NRQCD approach and found that a non-vanishing value of $B_S$ leads to a suppression of the cross-section in the small transverse-momentum region, with the suppression increasing as $B_S$ grows, expecting a better agreement with the experimental data.
Of course, this statement holds within the NP parametrization adopted in this work.

Looking ahead, our predictions for EIC kinematics highlight the significant potential of this future facility to disentangle the transverse dynamics of gluons in quarkonium production.
Precise low-$p_T$ measurements at the EIC will not only provide stringent tests of the TMD framework but also enable improved extractions of LDMEs in a kinematic region where standard NRQCD factorization fails.
Furthermore, EIC data will allow direct constraints on the non-perturbative parametrization, sharpening our understanding of soft-gluon dynamics and their interplay with heavy-quark hadronization.


\section*{Comments}

While this work was in preparation, a related study appeared in Ref.~\cite{Maxia:2025zee}.
It is important to stress that the two works differ in how the shape function has been described, as we explained in the main text.
Phenomenological predictions in Ref.~\cite{Maxia:2025zee} follow the matching between fixed order and TMD calculations, providing a crucial step toward a complete picture of the cross-section.
In contrast, and unlike this study, in our analysis we explicitly implement the calculation of the hard part at NLO.
Furthermore, our work focuses on the phenomenological implementation of the TMDShF described at the operator level in resummed cross-section predictions.
Our NNLL$'$ resummed predictions, supplemented by a consistent non-perturbative parametrization, yield narrower uncertainty bands and enhanced stability compared to NLL, underscoring the necessity of precision resummation and NP modeling in quarkonium TMD phenomenology.

\acknowledgments

We are grateful to Melih A. Ozcelik and Shubhendu P. Mandal for helpful discussions.
This work was supported by Ikerbasque (Basque Foundation for Science), by the Basque Government through the grant IT1628-22, and by the Spanish State Research Agency through the grants PCI2022-132984, PID2022-136510NB-C33 and CNS2022-135186.
\newpage
\appendix

\section{Notation and formulae}
\label{sec:appendix-1}
Some general constants and functions that appear in the main text include
\begin{equation}
    b_0 = 2 e^{- \gamma_E}, \quad  L_T = \ln ( \mu^2 b_T^2 e^{2 \gamma_E}/ 4), \quad N_\ve = \left( \frac{\mu^2 e^{\gamma_E}}{4 \pi} \right)^\ve  \; .
\end{equation}
where $\gamma_E=0.577$
\subsection*{TMD distributions}
The pure collinear matrix element for the incoming gluon is
\begin{equation}
\begin{aligned}
    J_n^{(0)\mu \nu} (x,\bm{k}_{n\perp}) & = \frac{x P^+}{2} \int \frac{d y^- d^2 \bm{y}_\perp}{(2 \pi)^3} e^{-i(xy^- P^+/2 - \bm{y}_\perp \cdot \bm{k}_{n \perp})} \\
    & \times \sum_{X_n} \sandwich{P}{ \mathcal{B}^{\mu,a}_{n\perp}(y^-,\bm{y}_\perp) }{X_n} \sandwich{X_n}{ \mathcal{B}^{\nu,a}_{n \perp}(0)}{P} \; ,
\end{aligned}
\end{equation}
such that the gluon TMD correlator is defined as $G_{g/N}^{\mu \nu} = J_n^{(0) \mu \nu} \sqrt{S}$, with $S$ the soft function, after zero-bin subtraction, and the regularization of the rapidity divergences.
In turn, the gluon correlator $G_{g/N}$ can be parametrized in terms of gluon TMDs, such that we have the following expression for an unpolarized $N$ nucleon:
\begin{equation}
    G_{g/N}^{\mu \nu}(x, k_T) = \frac{1}{2} \left( - g_T^{\mu \nu} f_1^g(x,k_T^2) + \frac{k_T^\mu k_T^\nu + \frac{1}{2} k_T^2 g_T^{\mu \nu}}{M^2} h_{1}^{\perp g} (x,k_T^2) \right) \; ,
\end{equation}
where the transverse projector is $g_T^{\mu \nu} = g^{\mu \nu} - P^\mu n^\nu - P^\nu n^\mu$ with $P^\mu$ the four-momentum of the nucleon, $n^\mu$ and $\bar n^\mu$ the four-vectors forming the light-cone basis and $M$ the nucleon mass.

The pure soft matrix elements for the outgoing $J/\psi$ bound-state are
\begin{equation}
\begin{aligned}
    S^{(0)}_{^1S_0^{[8]}} (\bm{k}_{s \perp}) & = \frac{1}{2}\frac{1}{N_c^2-1} \int \frac{d^2\bm{y}_\perp}{(2 \pi)^2} e^{i \bm{y}_\perp \cdot \bm{k}_{s\perp}} \\
    & \times \sum_{X_s} \sandwich{0}{\left[ \left( \mathcal S_v \mathcal S_n \right)^\dagger \chi^\dagger_{\bar{\bf{p}}} T^a \psi_{\bf{p}} \right](\bm{y}_\perp) \ket{J/\psi,X_s}  \bra{J/\psi,X_s} \left[  \mathcal S_v \mathcal S_n \psi^\dagger_{\bf{p}} T^a \chi_{\bar{\bf{p}}} \right](0)}{0} \; , \\
    S^{(0)}_{^3P_J^{[8]}} (\bm{k}_{s \perp}) & = \frac{\Delta_J^{i'j'ij}}{2 (2J+1) M^2 (N_c^2-1)} \sum_{\lambda} \int \frac{d^2\bm{y}_\perp}{(2 \pi)^2} e^{i \bm{y}_\perp \cdot \bm{k}_{s\perp}} \\
    & \times \sum_{X_s} \sandwich{0}{\left[ \left( \mathcal S_v \mathcal S_n \right)^\dagger \chi^\dagger_{\bar{\bf{p}}} (q^{i'} \sigma^{j'}) T^a \psi_{\bf{p}} \right](\bm{y}_\perp) \ket{J/\psi,X_s} \\
    & \times \bra{J/\psi,X_s} \left[  \mathcal S_v \mathcal S_n \psi^\dagger_{\bf{p}} (q^i \sigma^j) T^a \chi_{\bar{\bf{p}}} \right](0)}{0} \; ,
\end{aligned}
\end{equation}
such that the subtracted TMD shape function is defined as $S_{[n] \to J/\psi} = S^{(0)}_{[n] \to J/\psi} / \sqrt{S}$. The $\Delta_J$ are projectors onto the $J$-state:
\begin{equation}
\begin{aligned}
\Delta^{i'j' ij}_0  & = \frac{1}{3} \delta^{i'j'} \delta^{ij}  \; ,\\
\Delta^{i'j' ij}_1  & = \frac{1}{2} \left( \delta^{i i'} \delta^{j j'} - \delta^{i j'} \delta^{i' j} \right) \; ,\\
\Delta^{i'j'ij}_2  & = \frac{1}{2} \left( \delta^{i i'} \delta^{j j'} + \delta^{i j'} \delta^{i' j} \right) - \frac{1}{3} \delta^{i'j'} \delta^{ij} \; .
\end{aligned}
\end{equation}

\subsection*{Born cross-sections}

\begin{table}[t]
    \centering
    \begin{tabular}{|>{\columncolor{gray!15}}c|c|}
    \hline 
    $[n] = \,^1S_0^{[8]}$
    &
    $[1+(1-y)^2]$ \\
    \hline
    $[n] = \,^3P_0^{[8]}$
    &
        $ \frac{4}{3M^2} \left( \frac{\rho+3}{\rho+1} \right)^2 [1+(1-y)^2]$ \\
    \hline
    $[n] = \,^3P_1^{[8]}$
    &
        $\frac{8}{M^2} \frac{\rho(\rho+2)}{(\rho+1)^2} [1+(1-y)^2] - \frac{16 \rho}{M^2 (\rho+1)^2} y^2$ \\
    \hline
    $[n] = \,^3P_2^{[8]}$
    &
        $\frac{8}{3M^2} \frac{\rho^2 + 6 \rho + 6}{(\rho+1)^2} [1+(1-y)^2] - \frac{16\rho}{M^2 (\rho+1)^2} y^2$\\
    \hline
    \end{tabular}
    \caption{$\hat{\sigma}$ for electroproduction and for several spin configurations.}
    \label{tab:born_cross-sections}
\end{table}

The Born cross-sections for $J/\psi$ photo- and electroproduction can be obtained with the so-called projectors method (see, e.g., Ref.~\cite{Maltoni:1997pt}) or by using the matching tensors (see Eq.~(\ref{eq:Hadronic-Tensor})).
To present the results in a clean way, we write the Born cross-section as
\begin{equation}
    \sigma_{[n] \text{Born}} = \frac{4 \pi^2 \alpha_{em}^2 \alpha_s e_Q^2}{y \, Q^2 M (M^2 + Q^2)} \; \frac{\hat{\sigma}_{[n]}}{N_{pol}N_{col}}
\end{equation}
where the factors $\hat{\sigma}_{[n]}$ are gathered in the Table~\ref{tab:born_cross-sections}.
The parameter $N_{pol.}$ refers to the number of polarizations in $d=4-2\epsilon$ dimensions: $N_{J=0} = 1$, $N_1 = (3-2\epsilon)$ and $N_2 = (5-2\epsilon)(1-\epsilon)$. The parameter $N_{col}$ is the color normalization factor.
It was proven that these results coincide with those presented in Ref.~\cite{Bacchetta:2018ivt}.

\subsection*{Anomalous dimensions}

As mentioned in the text, the finite part of the TMD shape function anomalous dimension was obtained in Ref.~\cite{Echevarria:2024idp}:
\begin{equation}
    \gamma_s(\mu) = - \frac{\alpha_s(\mu)\, C_A}{\pi} \; .
\end{equation}

The evolution of the hard Wilson coefficient is given by
\begin{equation}
    \frac{d}{d \, \ln \mu} \, C_H(M^2,Q^2;\mu_H^2) = \left[ \Gamma_{cusp}(\alpha_s) \, \ln \left( \frac{f(M^2,Q^2)}{\mu_H^2} \right) + \gamma_{nc}(\alpha_s) \right] C_H(M^2,Q^2;\mu_H^2) \; ,
\end{equation}
so therefore, the anomalous dimension of the hard function $H = |C_H|^2$ is as follows:
\begin{equation}
\begin{gathered}
    \gamma_H = 2 \,  \Gamma_{cusp} (\alpha_s) \, \ln \left( \frac{f(M^2,Q^2)}{\mu_H^2} \right) + 2 \,  \gamma_{nc} (\alpha_s) \; , \\
    \text{with} \qquad \gamma_{nc} = - \frac{\alpha_s C_A}{2\pi} \; .
\end{gathered}
\end{equation}

\subsection*{Evolution of LDMEs}

After solving the renormalization-group-evolution equations for the LDMEs, one gets the following expression, e.g., see the Ref.~\cite{Butenschoen:2020mzi}:
\begin{equation}
\label{eq:LDME_evolution}
\begin{aligned}
    \braket{ ^1S_0^{[8]}}(\mu) & = \braket{ ^1S_0^{[8]} }(\mu_0) + \frac{16}{4 m^2} \left( \frac{1}{\beta_0} \ln \frac{\alpha_s(\mu_0)}{\alpha_s(\mu)} + \frac{\beta_1}{4 \pi \beta_0^2 } (\alpha_s(\mu) - \alpha_s(\mu_0)) \right)\\
    & \times \left[ \frac{C_F}{2 C_A} \braket{ ^1P_1^{[1]}} (\mu_0) + \frac{C_A^2 -4}{4 C_A} \braket{ ^1P_1^{[8]}} (\mu_0) \right] \; , \\
    \braket{ ^3P_J^{[8]}}(\mu) & = \braket{ ^3P_J^{[8]} }(\mu_0) + \frac{16}{4 m^2} \left( \frac{1}{\beta_0} \ln \frac{\alpha_s(\mu_0)}{\alpha_s(\mu)} + \frac{\beta_1}{4 \pi \beta_0^2 } (\alpha_s(\mu) - \alpha_s(\mu_0)) \right)\\
    & \times \left[ \frac{C_F}{2 C_A} \braket{ ^3D_{J+1}^{[1]}} (\mu_0) + \frac{C_A^2 -4}{4 C_A} \braket{ ^3D_{J+1}^{[8]}} (\mu_0) \right] \; ,
\end{aligned}
\end{equation}
where $\mu_0$ is the scale at which the values of the LDMEs mentioned throughout the text have been extracted, and $\mu$ is the scale at which we want to evolve the corresponding LDME. 
Therefore, given a state with quantum numbers $L$ and $J$, the evolution of the LDME is off diagonal depending on the LDME of the color-singlet and -octet states with $L+1$ and $J+1$.
Note that $\mu_0$ is normally equal to $m_T = \sqrt{M^2 + P_{\psi \perp}^2}$.

\subsection*{OPE matching coefficients}

As we mentioned in Eq.~\eqref{eq:OPE}, the TMDShF can be written in terms of their collinear analogous through the following matching coefficients, derived in Ref.~\cite{Echevarria:2024idp}:
\begin{equation}
\begin{aligned}
    C_{ ^1S_0^{[8]}}^S &= C_{ ^3P_J^{[8]}}^P = 1 + \frac{\alpha_s C_A}{2 \pi} L_T (1 - \ln \, \zeta_B) \; ,\\
    C_{ ^1P_1^{[1]}}^S & = C_{ ^3D_{J+1}^{[1]}}^P  = - \frac{\alpha_s }{2 \pi} \frac{8 C_F}{3m^2} L_T \; ,\\
    C_{ ^1P_1^{[8]}}^S & = C_{ ^3D_{J+1}^{[8]}}^P = - \frac{\alpha_s }{2 \pi} \frac{8 B_F}{3m^2} L_T \; ,
\end{aligned}
\end{equation}
where the superscript $S$ and $P$ are denoting the S-state and P-states LDMEs.

\section{One-loop virtual corrections} 
\label{sec:appendix-2}

As mentioned in the main text, in order to calculate the partonic cross-section for the process $\gamma^ + g \to c \bar{c}$ and thereby extract the NLO hard contribution, we have followed the procedure used for the case of a real photon, as presented in the Ref.~\cite{Maltoni:1997pt}.
In this appendix, we summarize the most relevant aspects of the computation and present partial results, aiming to provide the greatest possible level of detail.

The calculation of the cross-section is carried out in the non-relativistic limit of QCD, which refers to a regime where the velocities of the quarks involved in the strong interactions are much smaller than the speed of light.
For the production of a heavy-quark pair which will decay into a $J/\psi$, the momenta are typically decomposed as $p^\mu = P^\mu_\psi/2 + q^\mu$ and $\bar{p} ^\mu = P^\mu_\psi/2 - q^\mu$, where they are the momentum of the heavy quark and of the heavy anti-quark respectively.
The momentum $P_\psi$ is the momentum of the $J/\psi$ and  $q \sim m_c v$ is commonly referred to as the relative momentum, and refers to the typical momentum transferred between the heavy quarks when they are in the bound state.

The pair can be produced with different combinations of spin and angular momentum, which will determine the state they will form.
At the level of the process amplitude, the spin and angular momentum with which the pair is produced are determined by the Dirac matrices, and by the order in the expansion of $q$, respectively; as we can see below:
\begin{equation}
\begin{aligned}
\label{eq:def_amplitudes}
    \mathcal M^{\alpha \mu}_{0,0} & = \text{Tr} \left[ \mathcal C \, \Pi_0 \mathcal M^{\alpha \mu} \right]_{q=0} \; ,\\
    \mathcal M^{\alpha \mu}_{1,0} & = \text{Tr} \left[ \mathcal C \, \Pi_1^\rho \mathcal M^{\alpha \mu} \right]_{q=0} \varepsilon_\rho \; ,\\
    \mathcal M^{\alpha \mu}_{0,1} & = \frac{d}{dq_\sigma} \text{Tr} \left[ \mathcal C \, \Pi_0 \mathcal M^{\alpha \mu} \right]_{q=0} \varepsilon_\sigma \; ,\\
    \mathcal M^{\alpha \mu}_{1,1} & = \frac{d}{dq_\sigma} \text{Tr} \left[ \mathcal C \, \Pi_1^\rho \mathcal M^{\alpha \mu} \right]_{q=0} \mathcal \varepsilon_{\rho \sigma} \; .
\end{aligned}
\end{equation}
The notation used in the equation above is $\mathcal M_{S,L}$, where $\mathcal M$ is the QCD amplitude for the production via photon-gluon fusion without quark spinors, $\varepsilon_\sigma$ and $\varepsilon_{\rho \sigma}$ are polarization vectors and tensors describing the total angular momentum $J$, such that $|S-L| \leq J \leq S+L$.
Moreover, $\mathcal C = \sqrt{2} T_{ij}$ is the projector onto the color-octet state, and $\Pi_S$ are the projectors onto the spin states $S=0, 1$:
\begin{equation}
\begin{aligned}
    \Pi_0 & = \frac{1}{\sqrt{8m^3}} \left( \frac{\Slash P_\psi}{2} -  \Slash q - m \right) \gamma_5 \left( \frac{\Slash P_\psi}{2} +  \Slash q + m \right) \; , \\
    \Pi_1^\alpha & = \frac{1}{\sqrt{8m^3}} \left( \frac{\Slash P_\psi}{2} -  \Slash q - m \right) \gamma^\alpha \left( \frac{\Slash P_\psi}{2} +  \Slash q + m \right) \; .
\end{aligned}
\end{equation}
Note from Eq.~\eqref{eq:def_amplitudes} that after performing the trace and the derivative of the amplitude projected onto the corresponding spin and color state, the relative momentum is set to zero.
Thus, the scalar products, expressed in terms of the two hard scales of the process, $M = 2m$ and $Q^2$, take the following form:
\begin{equation}
\begin{gathered}
    q^2 = - Q^2, \quad p_g^2 = 0, \quad p^2 = \bar{p}^2 = m^2, \quad P_\psi^2  = 4m^2, \\
    q \cdot p = q \cdot \bar{p} = \frac{4m^2 - Q^2}{4} \quad \text{and} \quad p_g \cdot p = p_g \cdot \bar{p} = \frac{4m^2 + Q^2}{4} \; .
\end{gathered}
\end{equation}

The calculation is carried out in $d = 4 - 2\varepsilon$ dimensions, with renormalization performed in the $\overline{\text{MS}}$ scheme, implemented through the replacement $\mu^2 \to \mu^2 e^{\gamma_E}/(4\pi)$.
At NLO, one must also account for the renormalization of the heavy-quark mass.
In particular, this requires evaluating the contribution from the Feynman diagram generated by the mass counterterm $\delta m$, which amounts to computing the tree-level diagram with the insertion of the following Feynman rule on the internal heavy-quark line:
\begin{equation}
    - i N_\ve \delta m  = i \frac{\alpha_s C_F}{4 \pi} \left( \frac{3}{\veuv} + 3 \, \ln \frac{\mu^2}{m^2}  + 4 \right) m \; .
\end{equation}

Since there are nine independent Feynman diagrams and the contribution of each must be evaluated for the four states $,^1S_0^{[8]}$ and $,^3P_J^{[8]}$, presenting the full results would be excessively lengthy.
Therefore, beyond the points already discussed, we restrict ourselves to commenting on selected aspects of the calculation in the following.

\subsection*{Coulomb singularity}

We would like to clarify how we handle the Coulomb singularity in our calculation, because unlike the result for photoproduction~\cite{Maltoni:1997pt}, we do not explicitly obtain the Coulomb pole $1/\text{v}$ in Eq~\eqref{eq:virtual_corrections}. 

This singularity originates from the fact that, as the heavy quark–antiquark pair is produced at very short distances, the attractive Coulomb-like interaction drives the probability of them being at the same point to diverge.
In the NLO calculation, the Coulomb pole arises from the exchange of a gluon between the two heavy quarks.
More precisely, the pole emerges from the following type of one-loop integrals appearing in the evaluation of that diagram:
\begin{equation}
    I = \int \frac{d^d k}{(2 \pi)^d} \frac{1}{k^2 [(k-p)^2 -m^2] [(k+ \bar p)^2 -m^2]} \; ,
\end{equation}
where $k$ is the loop-momentum of the exchanged gluon.
Moreover, the integration in $k$ is carried out after projecting the amplitude onto the corresponding state and setting $q=0$, as we discussed in Eq.~\eqref{eq:def_amplitudes}, so actually we will need to calculate the integral for $\bar{p} = p$.
In fact, it is easy to see that the result of this integral after Partial fraction is the following:
\begin{equation}
\begin{aligned}
    \left. I \right|_{\bar{p} = p} & = -\frac{i 2^{2 \varepsilon -5} \pi ^{\varepsilon -2} m^{-2 (\varepsilon +1)} \Gamma (\varepsilon )}{2 \varepsilon +1}\\
    & = - \frac{i}{32 \pi^2 m^2} \left[ \frac{1}{\veuv} - \ln(m^2) - \gamma_E + \ln(4\pi) - 2 \right] + \mathcal O (\ve) \; .
\end{aligned}
\end{equation}
Note we do not obtain any dependence in the relative velocity v, as is obvious when setting $q = 0$.
However, the result reported in Ref.~\cite{Maltoni:1997pt} for this scalar integral contains an additional term inside the brackets of the previous expression, namely the Coulomb pole $-\pi^2/2\text{v}$.
This arises from considering $p \neq \bar{p}$ with $p \cdot \bar{p} = m^2 \bigl(\tfrac{1+\text{v}^2}{1-\text{v}^2}\bigr)$

Since the Coulomb divergence can ultimately be absorbed into the non-perturbative bound state, and given that we are consistent with the projector method defined in Eq.~\eqref{eq:def_amplitudes}, we shall not elaborate further on this difference.
Accordingly, all our results have been obtained by setting $p = \bar{p}$ in the evaluation of the one-loop integrals.

\subsection*{One-loop scalar integrals}

For completeness, in the following, we outline how we proceed with the calculation of the one-loop integrals and present the results for the master integrals we have used, including the expressions for the integrals denoted as $I_1$ and $I_2$ in Eq.~\eqref{eq:Finite-term}.

For a given diagram, our procedure consists of first reducing the propagators as much as possible by means of partial fractioning, while simultaneously performing a tensor decomposition of the numerator so as to rewrite it in terms of structures already present in the propagators.
For integrals involving propagators with powers larger than unity, we employ integration-by-part identities to simplify them further.
Ultimately, the amplitude is expressed in terms of a set of master integrals, i.e., those integrals which cannot be further reduced within this procedure.

Adopting the following notation for a master integral
\begin{equation}
    \text{MI}[\{p_1,m\},p_2,\{p_3,m\},\hdots] = N_\ve \int_k \frac{1}{[p_1^2-m^2]p_2^2[p_3^2-m^2]\hdots}\; ,
\end{equation}
we report the following results, which have been used in the calculation,
\begin{equation}
    \begin{gathered}
    \text{MI}[\{k,m\}]  = -i 2^{2 \varepsilon -4} \pi ^{\varepsilon -2} \Gamma (\varepsilon -1) \left(m^2\right)^{1-\varepsilon }    \\
    \text{MI}[\{ k,m \},k-p+q]  = -\frac{i 2^{2 \varepsilon -4} \pi ^{\varepsilon -2} \Gamma (\varepsilon ) m^{-2 \varepsilon }  }{\varepsilon -1} \, _2F_1\left(1,\varepsilon ;2-\varepsilon ;-\frac{Q^2}{2 m^2}-1\right)   \\
    \text{MI}[\{ k,m \}, k-p, \{ k-2p+q,m\}]  = -\frac{i}{48 \pi ^2 \left(4 m^2+Q^2\right)} \left(\pi ^2-6 \text{Li}_2\left(-\frac{Q^2}{2 m^2}-1\right)\right)   \\
    \text{MI}[\{ k,m \}, \{ k-2p+q,m \}]  = i (4 \pi )^{\varepsilon -2} \Gamma (\varepsilon ) \left(m^2\right)^{-\varepsilon }   \\
    \text{MI}[\{k,m\}^2,\{k-2p,m\}]  = - \frac{\pi - i 2}{64 \pi^2 m^2} \\
    \text{MI}[\{k,m\},\{k-q,m\}]  = - \frac{i \left( -Q(1+2\epsilon+\ln(\frac{\mu^2}{m^2}) + 2 \epsilon \sqrt{4m^2 + Q^2} \tanh^{-1} (\frac{Q}{\sqrt{4m^2+ Q^2}})) \right)}{16\pi^2 Q \epsilon}   \\
    \text{MI}[\{k,m\}, \{ k-2p,m \}, \{ k-q,m \}]  = \frac{i \left(\text{Li}_2\left(\frac{2 Q}{Q-\sqrt{4 m^2+Q^2}}\right)+\text{Li}_2\left(\frac{2 Q}{Q+\sqrt{4 m^2+Q^2}}\right)-\frac{\pi ^2}{2}\right)}{16 \pi ^2 \left(4 m^2+Q^2\right)}   \\
    \text{MI}[\{k, m\}^2,\{k-2p+q\}^2]  = \frac{i}{3}  2^{2 \varepsilon -5} \pi ^{\varepsilon -2} \Gamma (\varepsilon +2) m^{-2 (\varepsilon +2)} 
    \end{gathered}
\end{equation}
and the expressions for $I_1$ and $I_2$, in terms of the MIs,
\begin{equation}
\begin{aligned}
    I_1 & \equiv \text{MI}[\{k,m\},k-p,\{k-q,m\}] \; , \\
    I_2 & \equiv \text{MI}[\{k,m\}^2,k-p,\{k-2p+q,m\}^2] \; .
\end{aligned}
\end{equation}

\newpage

\subsection*{Analytical expressions for $f_{[n]}$}

We explicitly present below the expressions for the functions $f_{[n]}$, which enter the final result for the one-loop virtual corrections to the cross-section in Eq.~\eqref{eq:Finite-term}.

\begin{itemize}
\item \textbf{State $ ^1S_0^{[8]}$:}
\begin{flalign*} 
    f_1  & = C_A \frac{2  \rho }{2 \rho +1} - C_F\frac{4  \rho  (3 \rho +2)}{(2 \rho +1)^2}   \; , &  \\
    f_2  & = - C_A \frac{ (5 \rho +2)}{12 (\rho +1)} + C_F \frac{ (4 \rho +1)}{6 (\rho +1)}  \; ,\\
    f_3  & = C_A \frac{ \left(8 \rho ^2+13 \rho +3\right)}{3 \rho +1} - C_F \frac{2\left(16 \rho ^3+43 \rho ^2+28 \rho +5\right)}{(2 \rho +1) (3 \rho +1)}  \; ,\\
    f_4  & = C_A \frac{2  \rho }{2 \rho +1} - C_F\frac{4  \rho  (3 \rho +2)}{(2 \rho +1)^2}  \; ,\\
    f_5  & = \left(C_F - \frac{C_A}{2} \right)\frac{4 \rho  (5 \rho +1) }{\sqrt{\rho  (\rho +1)} (3 \rho +1)}  \; ,\\
    f_6 & = - C_A \frac{3  (2 \rho +1)}{2 (\rho +1)}+C_F \frac{ (2 \rho -1)}{\rho +1}  \; , \\
    f_7 & = - \left(C_F - \frac{C_A}{2} \right) \frac{2 \rho}{\rho +1}  \; ,\\
    f_8 & = \frac{\pi^2}{8}  \left( C_A (4 \rho +1) + 2 C_F (1-4 \rho ) \right) \;  ,\\
    f_9 & = f_{10} = 0  \; .
\end{flalign*}

\item \textbf{State $ ^3P_0^{[8]}$:}
\begin{flalign*}
    f_1 & = C_A \frac{2 \rho  \left(8 \rho ^2+9 \rho +3\right) }{(\rho +3) (2 \rho +1)^2} - C_F \frac{4 \rho  (3 \rho +2) \left(6 \rho ^3+19 \rho ^2+15 \rho +4\right)}{(\rho +1) (\rho +3) (2 \rho +1)^3}  \; , & \\
    f_2 & = C_A \frac{ \left(3 \rho ^3-36 \rho ^2-119 \rho -24\right)}{12 (\rho +1) (\rho +3)^2} - C_F \frac{ \left(4 \rho ^3-29 \rho ^2-104 \rho -15\right)}{6 (\rho +1) (\rho +3)^2} \; , & \\
    f_3 & = C_A\frac{ \left(24 \rho ^5+116 \rho ^4+133 \rho ^3+33 \rho ^2+13 \rho +9\right)}{(\rho +1) (\rho +3)^2 (2 \rho +1)} \\
    & - C_F \frac{2  \left(48 \rho ^6+256 \rho ^5+396 \rho ^4+269 \rho ^3+161 \rho ^2+89 \rho +21\right)}{(\rho +1) (\rho +3)^2 (2 \rho +1)^2} \; ,\\
    f_4 & = C_A \frac{2  \left(8 \rho ^4+17 \rho ^3+8 \rho ^2-\rho -1\right)}{(\rho +1) (\rho +3) (2 \rho +1)^2}\\
    & - C_F \frac{4  \left(18 \rho ^5+69 \rho ^4+75 \rho ^3+30 \rho ^2+2 \rho -1\right)}{(\rho +1) (\rho +3) (2 \rho +1)^3} \\
    f_5 & = C_F \frac{4  \rho ^3 (5 \rho +12)}{(\rho  (\rho +1))^{3/2} (\rho +3)}-C_A \frac{ \rho  \left(26 \rho ^4+109 \rho ^3+113 \rho ^2+3 \rho -3\right)}{2 (\rho  (\rho +1))^{3/2} (\rho +3)^2} \; ,\\
    f_6 & = - C_A \frac{\left(\rho ^3+19 \rho ^2+47 \rho +15\right)+2 C_F \left(\rho ^3-5 \rho ^2-17 \rho +3\right)}{(\rho +1) (\rho +3)^2} \; ,
    \end{flalign*}
\begin{flalign*}
    f_7 & = \left( C_F - \frac{C_A}{2}\right)  \frac{\left(2 \rho ^3-13 \rho ^2-46 \rho -3\right)}{ (\rho +1) (\rho +3)^2} \; ,\\
    f_8 & = \pi ^2 \left( C_A \frac{ \left(5 \rho ^2+13 \rho +3\right)}{2 (\rho +3)^2}- C_F \frac{ \left(10 \rho ^3+31 \rho ^2+14 \rho -3\right)}{2 (\rho +1) (\rho +3)^2}\right) \; ,\\
    f_9 & =  \left( C_F - \frac{C_A}{2}\right) \frac{ 2}{\rho ^2+4 \rho +3} \; , & \\
    f_{10} & = C_A \frac{3 \rho  \left(2 \rho ^4+3 \rho ^3-9 \rho ^2-19 \rho -1\right) }{2 (\rho  (\rho +1))^{3/2} (\rho +3)^2} + C_F \frac{12 \rho ^2 }{(\rho  (\rho +1))^{3/2} (\rho +3)}  \; .
\end{flalign*}

\item \textbf{State $ ^3P_1^{[8]}$:}
\begin{flalign*}
    f_1 & = - C_A\frac{2 \rho ^2-1}{(\rho -1) (2 \rho +1)}+2C_F \frac{2 \rho ^4+16 \rho ^3+9 \rho ^2-3 \rho -2}{(\rho -1) (\rho +1) (2 \rho +1)^2}  \; , &  \\
    f_2 & = - C_A \frac{5 \rho^2+13\rho+24}{12 (\rho +1)(\rho-1)} + C_F \frac{ (4 \rho^2+13\rho+25)}{6 (\rho +1)(\rho-1)} \; ,\\
    f_3 & = C_A\frac{\left(7 \rho ^2+22 \rho -1\right)}{4 (\rho -1) (\rho +1)}-C_F \frac{ \left(26 \rho ^3+47 \rho ^2-8 \rho -13\right)}{2 (\rho -1) (\rho +1) (2 \rho +1)} \; ,\\
    f_4 & = -C_A\frac{\left(2 \rho ^3+6 \rho ^2+\rho -1\right)}{(\rho -1) (\rho +1) (2 \rho +1)} + C_F\frac{2 \left(2 \rho ^4+24 \rho ^3+17 \rho ^2-\rho -2\right)}{(\rho -1) (\rho +1) (2 \rho +1)^2} \; ,\\
    f_5 & = \frac{C_A \left(-8 \rho ^3+23 \rho ^2+14 \rho +7\right)+16 C_F \left(\rho ^3-6 \rho ^2-2 \rho -2\right)}{4 \sqrt{\rho  (\rho +1)} \left(\rho ^2-1\right)} \; ,\\
    f_6 & = \frac{- C_A \left(6 \rho ^2+5 \rho +7\right) + 2 C_F \left(2 \rho ^2+5 \rho +11\right)}{2 \left(\rho ^2-1\right)} \; , \\
    f_7 & = \frac{\left(\rho ^2+3 \rho +6\right) (C_A-2 C_F)}{\rho ^2-1} \; ,\\
    f_8 & = \pi ^2 \left(\frac{C_A \left(4 \rho ^2+5 \rho +8\right)}{8 (\rho -1)}-\frac{C_F \left(4 \rho ^3+7 \rho ^2+17 \rho +10\right)}{4 (\rho -1) (\rho +1)}\right) \; ,\\
    f_9 & = -\frac{\rho  (C_A-2 C_F)}{(\rho -1) (\rho +1)} \; ,\\
    f_{10} & = -\frac{\sqrt{\frac{\rho }{\rho +1}} (6 \rho  (C_A-2 C_F))}{\rho ^2-1} \; .
\end{flalign*}

\item \textbf{State $ ^3P_2^{[8]}$:}
\begin{flalign*}
    f_1 & = - C_A\frac{\left(8 \rho ^4+6 \rho ^3-30 \rho ^2-27 \rho -6\right)}{(2 \rho +1)^2 \left(\rho ^2-3 \rho +6\right)}\\
    & + C_F\frac{2  \rho  \left(12 \rho ^5+78 \rho ^4-76 \rho ^3-249 \rho ^2-157 \rho -30\right)}{(\rho +1) (2 \rho +1)^3 \left(\rho ^2-3 \rho +6\right)}\; , \\
    f_2 & = \frac{C_A \left(3 \rho ^3-15 \rho ^2-26 \rho -12\right)+2 C_F \left(-4 \rho ^3+17 \rho ^2+23 \rho +6\right)}{12 (\rho +1) \left(\rho ^2-3 \rho +6\right)} \; ,\\
    f_3 & = C_A\frac{ \left(32 \rho ^5+34 \rho ^4+97 \rho ^3+60 \rho ^2+81 \rho +36\right)}{4 (\rho +1) (2 \rho +1) \left(\rho ^2-3 \rho +6\right)} \\
    & -\frac{C_F \left(64 \rho ^6+100 \rho ^5+116 \rho ^4+221 \rho ^3+546 \rho ^2+421 \rho +96\right)}{2 (\rho +1) (2 \rho +1)^2 \left(\rho ^2-3 \rho +6\right)} \; ,\\
    f_4 & = - C_A\frac{ \left(8 \rho ^5+14 \rho ^4-28 \rho ^3-25 \rho ^2+2 \rho +3\right)}{(\rho +1) (2 \rho +1)^2 \left(\rho ^2-3 \rho +6\right)} \\
    & + C_F \frac{2  \left(12 \rho ^6+78 \rho ^5-84 \rho ^4-189 \rho ^3-55 \rho ^2+23 \rho +9\right)}{(\rho +1) (2 \rho +1)^3 \left(\rho ^2-3 \rho +6\right)}\; ,\\
    f_5 & = -\frac{\rho  \left(C_A \left(4 \rho ^4-43 \rho ^3+124 \rho ^2+45 \rho -6\right)+8 C_F \rho  \left(2 \rho ^3+18 \rho ^2-51 \rho +3\right)\right)}{4 (\rho  (\rho +1))^{3/2} \left(\rho ^2-3 \rho +6\right)} \; ,\\
    f_6 & = \frac{2 C_F \left(-2 \rho ^3+4 \rho ^2+7 \rho +3\right)-C_A \left(2 \rho ^3-4 \rho ^2+19 \rho +27\right)}{2 (\rho +1) \left(\rho ^2-3 \rho +6\right)} \; , & \\
    f_7 & = -\frac{\left(2 \rho ^3-7 \rho ^2-10 \rho -3\right) (C_A-2 C_F)}{2 (\rho +1) \left(\rho ^2-3 \rho +6\right)} \; ,\\
    f_8 & = \frac{\pi ^2 \left(C_A \left(-4 \rho ^3+24 \rho ^2+55 \rho +27\right)+2 C_F \left(4 \rho ^3-32 \rho ^2-19 \rho -3\right)\right)}{8 (\rho +1) \left(\rho ^2-3 \rho +6\right)} \; ,\\
    f_9 & = \frac{(\rho -9) (C_A-2 C_F)}{2 (\rho +1) \left(\rho ^2-3 \rho +6\right)} \; ,\\
    f_{10} & = \frac{3 (\rho -9) \sqrt{\rho } (C_A-2 C_F)}{(\rho +1)^{3/2} \left(\rho ^2-3 \rho +6\right)} \; .
\end{flalign*}

\end{itemize}


\bibliographystyle{utphys}
\bibliography{references}

\end{document}